\documentclass[12pt]{article}
\pdfoutput=1
\usepackage{graphicx}
\usepackage{epstopdf}
\usepackage{amsmath}
\usepackage{amsfonts}
\usepackage{amssymb}
\usepackage{color}
\usepackage{mathrsfs}
\usepackage{hyperref}

\newcommand{\be}{\begin{equation}}
\newcommand{\ee}{\end{equation}}
\newcommand{\bea}{\begin{eqnarray}}
\newcommand{\eea}{\end{eqnarray}}
\newcommand{\barr}{\begin{array}}
\newcommand{\earr}{\end{array}}
\newcommand{\la}{\left\langle}
\newcommand{\ra}{\right\rangle}
\newcommand{\dqc}{\frac{d^3q}{(2\pi)^3}}
\newcommand{\dqcp}{\frac{d^3q'}{(2\pi)^3}}

\newcommand{\derd}{d}

\newcommand{\omnot}{\Omega_{m,0}}
\newcommand{\deltal}{\delta_l}

\newcommand{\dkc}{\frac{\derd^3k}{(2\pi)^3}}

\def\beq{\begin{equation}}
\def\eeq{\end{equation}}
\def\be{\begin{equation}}
\def\ee{\end{equation}}
\def\bea{\begin{eqnarray}}
\def\eea{\end{eqnarray}}
\def\d{{\partial}}

\begin{document}


\setcounter{page}{1} \baselineskip=15.5pt \thispagestyle{empty}


\begin{center}

{\Large \bf Galaxy Bias and non-Linear Structure Formation in General
Relativity} \\[0.4cm]
{\large Tobias Baldauf$\,{}^{a}$, Uro\v s Seljak$\,{}^{a,b,c}$,\\[0.3cm]
 Leonardo Senatore$\,{}^{d,e}$  and Matias Zaldarriaga$\,{}^{f}$}
\\[0.5cm]
{\normalsize {\sl (a) Institute for Theoretical Physics, University of Zurich, Zurich, Switzerland}}\\ \vspace{.2cm}
{\normalsize {\sl (b) Physics Department, Astronomy Department and Lawrence Berkeley National Laboratory, University of California, Berkeley, CA, USA}}\\ \vspace{.2cm}
{\normalsize {\sl (c) Institute for the Early Universe, EWHA Womans University, Seoul, South Korea}}\\ \vspace{.2cm}
{\normalsize {\sl (d) Stanford Institute for Theoretical Physics, Stanford University, Stanford, CA, USA}}\\ \vspace{.2cm}
{\normalsize {\sl (e) Kavli Institute for Particle Astrophysics and Cosmology, Stanford, CA, USA}}\\ \vspace{.2cm}
{\normalsize {\sl (f) School of Natural Sciences, Institute for Advanced Study, \\Olden Lane, 
Princeton, NJ 08540, USA}}\\
\vspace{.2cm}

\end{center}


\hrule \vspace{0.3cm}
{\small  \noindent \textbf{Abstract} \\[0.3cm]
\noindent
Length scales probed by the large scale structure surveys are becoming closer
and closer to the horizon scale. Further, it has been recently understood that
non-Gaussianity in the initial conditions could show up in a scale dependence of
the bias of galaxies at the largest possible distances. It is therefore
important to take General Relativistic effects into account. Here we provide
a General Relativistic generalization of the bias that is valid both for
Gaussian and for non-Gaussian initial conditions. The collapse of objects
happens on very small scales, while long-wavelength modes are always in the
quasi linear regime. Around every small collapsing region, it is therefore
possible to find a reference frame  that is valid for arbitrary times and where
the space time is almost flat: the Fermi frame. Here  the Newtonian
approximation is applicable and the equations of motion are the ones of the
standard $N$-body codes. The effects of long-wavelength modes are encoded in the
mapping from the cosmological frame to the local Fermi frame. At the level of
the linear bias, the effect of the long-wavelength modes on the dynamics of the
short scales is all encoded in the local curvature of the Universe, which allows
us to define a General Relativistic generalization of the bias in the standard
Newtonian setting. We show that the bias due to this effect goes to
zero as the square of the ratio between the physical wavenumber and the Hubble
scale for modes longer than the horizon, confirming the intuitive picture that
modes longer than the horizon do not have any dynamical effect. On the other
hand, the bias due to non-Gaussianities does not need to vanish for modes longer
than the Hubble scale, and for non-Gaussianities of the local kind it goes to a
constant. As a further application of our setup, we show that it is not
necessary to perform large $N$-body simulations to extract information about
long-wavelength modes: $N$-body simulations can be done on small scales and
long-wavelength modes are encoded simply by adding curvature to the
simulation, as well as rescaling the time and the scale. 

}
 \vspace{0.3cm}
\hrule
\vfil



\section{Introduction and Summary\label{sec:introduction}}

Large Scale Structure (LSS) surveys are becoming larger and larger, and soon
they will be able to probe cosmological modes whose length scale is comparable
to the Hubble scale. General Relativistic effects scale as the ratio of the
physical wavenumber $k/a$ and the Hubble scale
\be
{\rm General\ Relativistic \ Effects }\sim \left(\frac{H a}{k}\right)^2\ ,
\ee
and it is therefore important to take these effects into account in order to be
able to interpret next generation of LSS data. All the relativistic effects are
basically projection effects relating what happens in one place to what we see:
they include such things as  lensing, redshift, distortion, gravitational
redshift, etc. A consistent derivation of them for dark matter has been recently
performed in~\cite{Yoo:2009au}. Unfortunately we do not observe dark matter
directly, but just luminous objects. From the observation of them we are able to
reconstruct the dark matter density field by the realization that collapsed
objects are biased tracers of the dark matter field.  The concept of bias has so
far always been defined using the Newtonian approximation that is valid for
small length scales. The purpose of this paper is to provide a generalization of
this concept that is valid at arbitrary long-wavelengths.

Another reason that motivates us to provide such a generalization is due to
the recent observation that non-Gaussianity in the primordial density field can
induce a scale dependence in the bias at large
wavelengths~\cite{Dalal:2007cu,Slosar:2008hx}. In the presence of
non-Gaussianities of the local kind, the bias receives a scale dependence that
in the Newtonian treatment behaves as
\be\label{eq:intro-bias}
\delta_{n_g}(k)=b(k)\delta_m(k)\ , \qquad b_{f_{\rm NL}^{loc.}}\sim b_{f_{\rm NL}^{loc.}=0}\left(1+f_{\rm NL}^{\rm loc.}\frac{H^2 a^2}{k^2}\right) \ ,
\ee
where $\delta_{n_g}$ is the perturbation to the density of objects, $\delta_m$
the perturbation to the matter density, and $k$ is the wavenumber of the mode,
and where we have neglected factors of order unity and the transfer function for
simplicity. The important point of this expression is that in the presence of
non-Gaussianities that have a non-vanishing squeezed limit, such as the ones of
the local kind or the new ones that have been found in the Effective Field
Theory of Multifield Inflation~\cite{Senatore:2010wk} with support both on equilateral and squeezed configurations, the bias receives a scale
dependence at large scales  proportional to $f_{\rm NL}$. This provides an ideal
setup for measuring non-Gaussianities in LSS, as the signal is peaked on large
scales, where theoretical predictions are under better control. Indeed current
limits on $f_{\rm NL}^{\rm loc.}$ obtained from the Sloan Digital Sky Survey
(SDSS)
data are already competitive with the ones from WMAP~\cite{Slosar:2008hx}, and analysis of the bispectrum is expected to be even more promising~\cite{Baldauf:2010vn}. 

An odd feature of (\ref{eq:intro-bias}) is that
\be
b_{f_{\rm NL}^{loc.}} \rightarrow \infty \quad {\rm as}\quad k\rightarrow 0\ .
\ee
It is equally strange that the standard Gaussian bias does not go to zero as $k\rightarrow0$: one might indeed expect that modes longer than the Hubble scale should have no effects on the local dynamics. Of course, all of these results are due to the fact that we are trusting (\ref{eq:intro-bias}) way into a regime where it does not apply: as $k/a$ becomes close to $H$, a proper General Relativistic treatment becomes necessary.

The main purpose of this paper is to provide such a General Relativistic
generalization of the bias that is valid both in the case of primordial Gaussian
and non-Gaussian initial conditions. In doing this, we will also provide a way
to understand small $N$-body simulations in the General Relativistic setting,
and to show that in order to study the effects of long-wavelength modes, it is
not necessary to run large, time consuming, $N$-body simulations. Let us briefly
summarize the logic and the main results.
\begin{itemize}

\item Cosmological perturbations become non-linear and lead to collapse only on
very small scales, where the Newtonian approximation is valid. This suggests
that if we insist on describing length scales much smaller than the Hubble
scale, then the current Newtonian description is valid.

\item Given a perturbed Friedman Robertson Walker (FRW) Universe with
fluctuations of arbitrary length scale, it is possible to identify a coordinate
frame valid on spatial  distances much smaller than the horizon and for an
arbitrary amount of time, where the metric appears locally as the one of
Minkowski space, with small perturbations of order $(H x)^2$, $x$ being the
spatial distance from the origin. These coordinates represent the inertial
frame of a free falling observer, and they are called Fermi
coordinates~\cite{MM}. In the case where the matter is non-relativistic, in this
frame the Newtonian approximation is manifest, and we argue that this is the
frame where results of small-box $N$-body simulations can be interpreted. We
explicitly construct such a reference frame at linear order in the long scale
fluctuations for a spherically symmetric configuration of the long-wavelength
modes, as this is sufficient for the description of linear bias.
Generalizations to different configurations for the long-wavelength modes or to
the non-linear level should be straightforward.

\item In these coordinates, all the effect of the long-wavelength mode is
included in the mapping from the global frame to the Fermi frame, and in the
long-wavelength curvature of the local patch. Since for the linear bias we can
use spherical symmetry for the long-wavelength modes, the long-wavelength part
of the Fermi metric must be equivalent to that of a curved FRW Universe, and
therefore all the effect that a long-wavelength mode has on the local dynamics
is indeed in the curvature of the local FRW Universe. This is given by
\be
\Omega_K\sim \frac{\nabla^2\zeta(\vec x_{L}, t_{L})}{a^2 H^2}\ ,
\ee
where $\zeta$ is the curvature perturbation in comoving gauge, and here for
simplicity we have omitted numerical factors given later in the text.

\item This allows us to generalize the concept of bias to the General
Relativistic setting, by declaring it to be the derivative of the proper number
density of objects at a fixed proper time with respect to the curvature of the
local Universe:
\be
b\sim\frac{1}{n_p}\frac{\d n_p}{\d\Omega_K}\  \quad\Rightarrow \quad \delta_{n_p}\sim b \frac{\nabla^2\zeta}{a^2 H^2}+\ldots\ ,
\ee
where $n_p$ is the proper number density of objects, $\delta_{n_p}$ their
relative overdensity and the dots stand for additional terms coming from various
projection effects that we will discuss in the text. Here we have neglected
numerical factors. This expression makes sense physically, as for modes much
longer than the Hubble scale, $\Omega_K\rightarrow 0$, making explicit the
General Relativistic statement that metric modes that have no measurable
gradients do not affect the local dynamics.

\item In presence of primordial non-Gaussianities, the initial conditions for
the fluctuations in the Fermi patch can depend on other parameters. In the case
of non-Gaussianities of the local kind, initial conditions depend explicitly on
$\zeta$, a quantity that has no effect on the local dynamics. In this case, we
extend the definition of the bias to include the derivative of the proper number
density of objects with respect to the parameter itself. For example, in the
case of non-Gaussianities of the local kind, we have
\be
b_{f_{\rm NL}^{loc.}}\sim\frac{1}{n_p}\frac{\d n_p}{\d\zeta}\quad\Rightarrow
\quad \delta_{n_p}\sim b \frac{\nabla^2\zeta}{a^2 H^2}+b_{f_{\rm NL}^{loc.}}
\zeta + \ldots\ ,
\ee 
where again the $\ldots$ stand for additional terms coming from various projection effects which we will discuss in the text. 
We see that the relative factor of $k^2$ between the standard bias and the one
induced by $f_{\rm NL}$ is preserved in the General Relativistic limit. However,
most importantly, the physical effect of long-wavelength fluctuations on the
local overdensity does not blow up as $k\rightarrow 0$: it is simply the fact
that the standard Gaussian effect goes to zero while the non-Gaussian one stays
constant.

\item Finally we point out that our construction of the local Fermi coordinates
shows that it is not strictly required to run time-consuming large-box $N$-body
simulations to study the effect of long-wavelength fluctuations: their effect
can be simply included by running small-box $N$-body simulations with different
cosmological parameters than in the standard cosmology.

\end{itemize}

Related works on the way to include long-wavelength perturbations inside
small-box $N$-body simulations have appeared
in~\cite{Tormen1995,Cole1996,Schneider2011}. Related work on the way to
derive
the bias of the local form in the General Relativistic context has appeared
in~\cite{Sasaki:1987ad,Dodelson:2008qc,Wands:2009ex,Yoo:2010ni}
and a connection of the latter to primordial non-Gaussianities of the local
form has been made in \cite{Bartolo:2010ec}.

\section{Fermi Coordinates for Perturbed FRW\label{sec:FermiCoordinates}}

Given a sufficiently smooth spacetime, it is possible to identify a set of coordinates centered around a timelike geodesic, known as Fermi coordinates~\cite{MM}. They have two important properties: the metric is approximately that of Minkowski space, with corrections that start quadratically in the (space-like) geodesic distance from the time-like geodesic taken as the origin, and they are valid in the (spatial) vicinity of the time-like geodesic for all times. 

In an FRW spacetime the Hubble expansion appears in the Fermi coordinates as
a small correction to the standard dynamics in Minkowski space. This set of
coordinates was found for unperturbed FRW first in \cite{Cooperstock:1998ny}.
Here we are going to provide such a set of coordinates for a linearly perturbed
FRW Universe. We will then argue that in this set of coordinates the Newtonian
approximation is valid, and that this is actually the frame in which $N$-body
simulations are performed. Furthermore, we will provide a mapping
from the local Fermi coordinates to the global
coordinates of a perturbed FRW, and we will show how simulations have to
be performed in order to include the effect of perturbations with wavelengths 
larger than the box size.

Let us therefore find these coordinates. Let us suppose we have an FRW metric
with some linear long-wavelength fluctuations. We
start from a perturbed FRW metric in Newtonian gauge:
\be
ds^2=-\left(1+2\Phi(\vec x_G, t_G)\right)dt^2_G+a(t_G)^2\left(1-2\Psi(\vec x_G, t_G)\right)d\vec x_G^2\ .
\ee
In app.~\ref{app:Fermi-zeta} we perform the same construction starting from $\zeta$-gauge.
Here the subscript $_G$ stands for Global to stress that these coordinates are
valid for the entire FRW space. A great simplification comes from the fact that
we wish to study the properties of the large scale structures mainly in the
regime where the long-wavelength modes are linear: in other words, we are mainly
interested in the two-point function of large-scale fluctuations. This has two
consequences. First,  the behavior of $\Phi$ and $\Psi$ can be found by solving
the linear Einstein equations and the linearized equations of motion for matter.
For example, we can assume that there is no anisotropic stress at linear level,
so that $\Psi=\Phi$. Second, if we wish to compute scalar quantities (as we will
wish), we can use superposition principle to restrict ourselves to consider
configurations where $\Phi$ is spherically symmetric around one point, let us
say the point $\vec x_G=0$. Generalization to the non-linear treatment of $\Phi$
is conceptually straightforward, but computationally not so, and we leave it to
future work~\footnote{Of course such a non-linear treatment would become much
more pressing if we had convincing evidence that the primordial perturbation
were non-Gaussian. There is some reason of possible excitement: in the CMB
Gaussianity is excluded only at the $2\sigma$ level \cite{Komatsu:2010fb}
through the analysis of the three-point function of the orthogonal kind
parametrized by $f_{\rm NL}^{\rm orthog.}$~\cite{Senatore:2009gt}.}.

 \begin{figure}[h!]
    \centering
        \includegraphics[width=.60\textwidth]{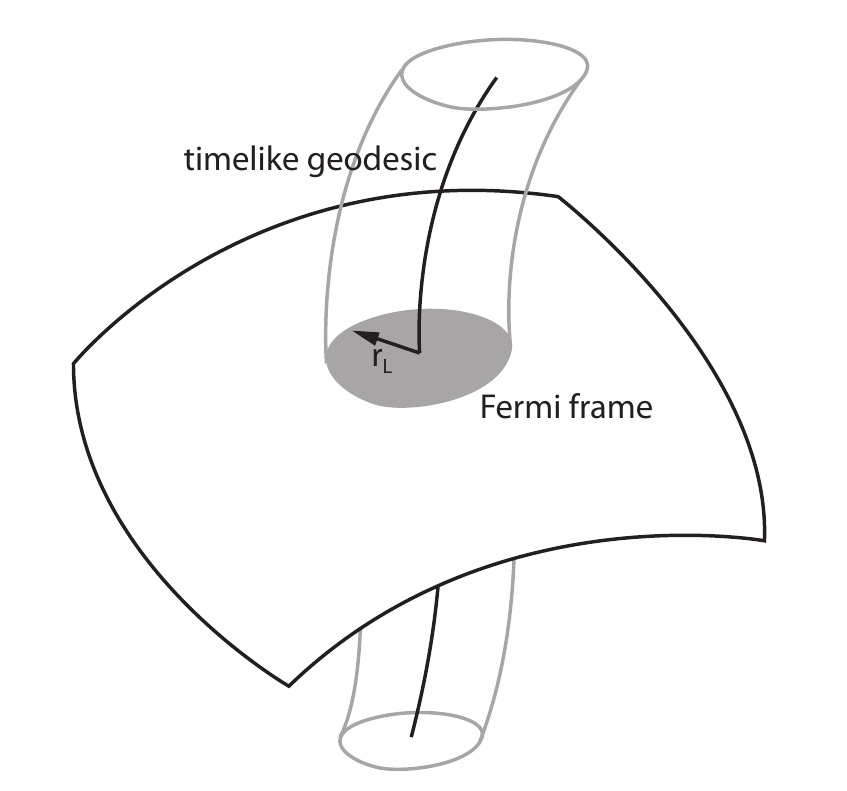}
    \caption{\sl Fermi Coordinates.}
    \label{fig:localpatch}
\end{figure}

In order to find the Fermi coordinates (fig.~\ref{fig:localpatch}), we can
restrict ourselves to the neighborhood of a time-like geodesic. Spherical
symmetry suggests to consider the geodesic $\vec x_G(t_G)=0$. If we consider
modes whose wavelength is much larger than the region of interest, we can Taylor
expand the metric around the origin, and keep only the leading two derivatives.
Notice that numerical simulations have to follow dark matter particles, and
therefore their region of interest corresponds to scales corresponding to the
length traveled by the particles, of the order of the non-linear scale. We
obtain:
\be
ds^2\simeq-\left(1+2\Phi(\vec 0, t_G)+\Phi(\vec 0,t_G)_{,r_Gr_G} r_G^2\right)dt^2_G+a(t_G)^2\left(1-2\Phi(\vec 0, t_G)-\Phi(\vec 0, t_G)_{,r_Gr_G} r_G^2\right)d\vec x_G^2\ ,
\ee
where $r_G^2=x_{G, 1}^2+x_{G, 2}^2+x_{G, 3}^2$.
We can find the coordinates in which the above metric appears in the Fermi way
in a simple, but brute force, way that we describe here. A more geometric
derivation is presented in app.~\ref{app:geometrical}. Let us first warm up by
considering the case of an unperturbed, curved FRW Universe, whose metric is of
the form
\be
ds^2=-dt_G^2+a(t_G)^2\frac{d\vec x_G^2}{\left[1+\frac{1}{4}K\,\vec x_G^2\right]^2}\ .
\ee
We consider the curved case here because it will be useful for later purposes. It is easy to check that upon the following change of coordinates, valid at small distances~\cite{Cooperstock:1998ny}:
\bea
&&t_G= t_L-\frac{1}{2}H(t_L)r_L^2\ , \\ \nonumber
&& x_G^i =\frac{x_L^i}{a(t_L)}\left(1+\frac{1}{4}H(t_L)^2r_L^2\right)\ ,
\eea
where $r_L^2=x_{L, 1}^2+x_{L, 2}^2+x_{L, 3}^2$ and the subscript $_L$ reminds us that these are the Locally valid coordinates, the metric takes the form
\bea\label{eq:FermiFRW}
ds^2&=& -\left[1-\left(\dot
H(t_L)+H(t_L)^2\right)r_L^2\right]dt_L^2+\left[1-\frac{1}{2}\left(H(t_L)^2+\frac
{K}{a(t_L)^2}\right)r_L^2\right]d\vec x_L^2 \ .
\eea
As we had anticipated, for an indefinite amount of time, the metric near the spatial origin is approximately the Minkowski one, with corrections starting at order $r_L^2$ and suppressed by powers of $H\, r_L\ll1 $. So for example this metric is valid for distances smaller than Hubble, but it clearly can include cosmologically interesting length scales such as the non-linear scale where structures form.

To consider now the generic perturbed FRW flat space, let us generalize the change of coordinates as
\bea\label{eq:change}
&&t_G= t_L-\frac{1}{2}H(t_L)r_L^2-\int_0^{t_L} \Phi(\vec 0,t')dt'+g_1(t_L) r_L^2 \ , \\ \nonumber
&& x_G^i =\frac{x_L^i}{a(t_L)}\left(1+\frac{1}{4}H(t_L)^2r_L^2+f_1(t_L)+f_2(t_L)r_L^2\right)\ ,
\eea
and let us determine the functions $f_{1,2},\ g_1$, meant to be first order in the metric fluctuations, by imposing that the metric in the local coordinates is of the Fermi form, with the additional constraint that the spatial part be proportional to $\delta_{ij}$. Notice that we have made the educated guess that at the origin the Local time equals the proper time. We will verify shortly that this is a good guess. After some straightforward algebra, we obtain
\bea\nonumber\label{eq:LocalGlobalMapping}
t_G&=& t_L-\int_0^{t_L} \Phi(\vec 0,t_L)dt'\\ \nonumber 
&&-\left(\frac{1}{2}H(t_L)-H(t_L)\Phi(\vec 0,t_L)-\frac{1}{2}\Phi(\vec 0,t_L)_{,t_L} -\frac{\dot H(t_L)}{2}\int_0^{t_L}\Phi(\vec 0,t')dt'\right) r_L^2 \ , \\ \nonumber
x_G^i &=&\frac{x_L^i}{a(t_f)}\left[1+\Phi(\vec 0,t_L)+H(t_L)\int_0^{t_L}\Phi(\vec 0, t')dt'+\right.\\ \nonumber &&\!\!\!\!\!\!\!\!\!\!\!\!\!\!\!\!\!\!\!\!\frac{1}{4}\left.\left(H(t_L)^2+H(t_L)\left(H(t_L)^2-2\dot H(t_L)\right)\int^{t_L}_0\Phi(\vec 0,t')dt'-H(t_L)^2\Phi(\vec 0,t_L)-2 H(t_L)\Phi(\vec 0,t_L)_{,t_L}\right)r_L^2\right]\ .\\
\eea
Let us recall the common definition of the comoving-gauge curvature perturbation $\zeta$
\be
\zeta(\vec x_G,t_G)=-\Phi(\vec x_G,t_G)+\frac{H(t_G)^2}{\dot
H(t_G)}\left(\Phi(\vec x_G,t_G)+\frac{\dot\Phi(\vec x_G,t_G)}{H(t_G)}\right)\ ,
\ee
and the fact that this is constant for adiabatic fluctuations and for
wavelengths longer than the sound horizon:
\be
\dot\zeta(\vec x_G,t_G)=\frac{H(t)}{\dot H(t)}\left[\ddot \Phi(t)
+\left(H(t)-\frac{\ddot H(t)}{\dot H(t)}\right)\dot\Phi(t)+\left(2\dot
H(t)-\frac{\ddot H(t) H(t)}{\dot H(t)}\right)\Phi(t)\right]=0\ ,
\ee
where the dot stays for derivative with respect to the time variable.
This implies that we can write $\zeta$ as
\be
\zeta(t)=-\Phi(t)-H(t)\int_0^tdt'\, \Phi(t')\ ,  \quad
\Rightarrow\quad\dot\Phi(t)+ H(t)\Phi(t)+\dot H(t)\int_0^t dt' \Phi(t')=0\ ,
\ee
and therefore we can simplify the former expressions to get
\bea\nonumber\label{eq:LocalGlobalMappingSimpl}
t_G&=& t_L-\int_0^{t_L} \Phi(\vec 0,t_L)dt'-\frac{1}{2}H(t_L)\left(1-\Phi(\vec 0,t_L)\right) r_L^2 \ , \\ 
x_G^i &=&\frac{x_L^i}{a(t_f)}\left[1+\frac{H(t_L)^2}{4}
r_L^2\right]\left(1-\zeta(\vec 0,t_L)\right)\ .
\eea
 The resulting metric is of the form
\bea\nonumber\label{eq:metricPerturbedFermi}
ds^2&=& -\left[1-\left\{\dot H(t_L)+H(t_L)^2-2\left(H(t_L)^2+\dot H(t_L)\right)\Phi(\vec 0, t_L)-3H(t_L)\Phi(\vec 0, t_L)_{,t_L}\right.\right.\\ \nonumber &&\quad \!\!\!\!\!\left.\left.-\Phi(\vec 0, t_L)_{,t_Lt_L}-\left(2H(t_L) \dot H(t_L)+\ddot H(t_L)\right)\int_0^{t_L}\Phi(\vec 0,t')dt'-\frac{\Phi(\vec 0,t_L)_{,r_G r_G}}{a(t_L)^2}\right\}r_L^2\right]dt_L^2+\\ \nonumber
&&+\left[1-\left\{\frac{H(t_L)^2}{2}-H(t_L)^2\Phi(\vec 0,t_L)-H(t_L)\Phi(\vec 0,t_L)_{,t_L}
\right.\right.\\ &&\quad\left.\left.
-H(t_L)\dot H(t_L)\int_0^{t_L}\Phi(\vec 0, t')dt'+\frac{\Phi(\vec 0,t_L)_{,r_G r_G}}{a(t_L)^2}\right\}r_L^2\right]d\vec x_L^2 \ .
\eea
which is valid without assuming that $\zeta$ is constant. If we use that $\zeta$ is indeed constant outside of the sound horizon, the metric simplifies to
\bea\label{eq:metricPerturbedFermiSimpl}
&&ds^2=-\left[1-\left(\dot H(t_L)+H(t_L)^2-\frac{\Phi(\vec 0,t_L)_{,r_G r_G}}{a(t_L)^2}\right)r_L^2\right]dt_L^2\\ \nonumber
&&\qquad\ \  \,+\left[1-\left(\frac{H(t_L)^2}{2}+\frac{\Phi(\vec 0,t_L)_{,r_G r_G}}{a(t_L)^2}\right)r_L^2\right]d\vec x_L^2 \ . 
\eea
The above metric represents the description of a perturbed FRW Universe on
scales much smaller than the typical length scale of the perturbations. For this
reason, in the presence of adiabatic perturbations whose wavelength is longer
than the sound horizon, it has to be equivalent to the local version of an FRW
metric, as represented in the local coordinates of~(\ref{eq:FermiFRW}). This is
indeed due to Birkhoff theorem. This is in fact true: upon identification of an
effective local expansion rate $H_L(t_L)$ and of an effective curvature $K_L$
given by
\bea\label{eq:localvariables}
&&H_L(t_L)=H(t_L)+\frac{1}{ H(t_L)a(t_L)^2}\left(\Phi(\vec 0,t_L)+\zeta(\vec
0,t_L)\right)_{,r_G r_G}\ , \\ \nonumber
&&K_L=2 \left[\Phi(\vec 0,t_L)-\frac{H(t_L)^2}{\dot H(t_L)}\left(\Phi(\vec
0,t_L)+\frac{\Phi(\vec 0,t_L)_{,t_L}}{H(t_L)}\right)\right]_{,r_G
r_G}=-\frac{2}{3}\nabla^2_G\zeta(\vec 0,t_G)\ ,
\eea
where $H_L(t_L)=\dot a_L(t_L)/a_L(t_L)$, the metric
(\ref{eq:metricPerturbedFermi}) takes the form of the curved unperturbed FRW
Universe in (\ref{eq:FermiFRW}) with the simple replacement $a \rightarrow a_L,\
K\rightarrow K_L$~\footnote{It might be useful to notice that the curvature
perturbation $K=-\frac{2}{3}\nabla^2\zeta$ can be expressed in terms of the
matter  density perturbation in comoving gauge $\delta^{(com)}_l$ as
\be
K=\left(\Omega_{m,0}+\frac{2}{3}f_0\right)H_0^2 
\delta^{(com)}_{l,0} \ ,
\ee
where $H_0$ is the Hubble parameter at the present time, $\Omega_{m,0}$ is
the fraction of energy in matter at present time, and
$f=\frac{\d \log D}{\d \log a}$ with $D$ being the growth factor such that
$\delta^{(com)}(t)=D(t)\delta^{(com)}_0$.The subscript 0 is used for quantities
evaluated at redshift zero.
}. 
In this case, the local curvature $K_L$ is proportional to the Laplacian
of the
curvature perturbation usually denoted by $\zeta$, and is thus constant in time.
$H_L$ follows the normal Friedmann equations for a curved FRW.

In summary, we have been able to see that an FRW Universe with a linear
adiabatic perturbation whose wavelength is longer than the sound horizon can be
described, locally, by a metric that is very close to the Minkowski one, and is
actually equivalent to one of a curved FRW Universe. The assumption of
adiabaticity and that the wavelength of the mode is longer than the sound
horizon is necessary in order for the curvature of the Universe to be constant
in time: it is only in this case that there is one single local history for the
Universe, which implies that the long-wavelength mode at linear level can be
completely re-absorbed into the curvature of a local FRW Universe. In practice,
this implies that our method of dealing with long wavelength perturbations is
applicable to adiabatic long-wavelength fluctuations in the case where the speed
of sound of the fluctuations is very small. This includes a Universe filled with
dark matter and a cosmological constant, or with dark matter and quintessence
with a very small speed of sound as the models studied in
\cite{Creminelli:2006xe}, while it does not apply to models with quintessence
with non-vanishingly small speed of sound.

Although for some questions one can restrict to the case of spherical symmetry,
this is not possible in general. The Fermi coordinates exist also in the absence
of spherical symmetry. In app.~\ref{app:Fermi-wave} we present the form of the
Fermi coordinates starting in Newtonian gauge with a plane wave perturbation.

\subsection{A Simple Check}
It is worth to show explicitly how our procedure works in a practical example,
where the long-wavelength fluctuation is short enough to allow for a Newtonian
treatment. Since we just said that the effect of a long mode can be re-absorbed
in a curvature of the background (at linear level and after using superposition
principle), this suggests that we should be able to re-derive the growth
function at second order for short wavelength fluctuations in the presence of
longer, spherically symmetric fluctuations as derived
in~\cite{Bernardeau:2001qr} in the standard perturbation theory approach. Here
instead we derive it from the growth of modes in a curved Universe. Working only
in the limit where all the modes are describable within the Newtonian
approximation and working in Einstein-de-Sitter space, this calculation is
carried out in app.~\ref{app:collobjects}, and here we summarize the main
results.

The evolution of short modes in the effective curved Universe is related to the short modes in an flat Universe $\delta_\text{s,flat}$ as
\be
\delta_s(\vec x)=\delta_{s,\rm{flat}}(\vec x)\left(1+\frac{34}{21}\delta_l(\vec x)\right)\ ,
\ee
where we have restricted ourselves to the matter only Einstein-de-Sitter
Universe. In Fourier space we get
\be
\delta_s(\vec
k_s)=\delta_{s,\rm{flat}}(\vec
k_s)\left(1+\frac{34}{21}\delta_l(
k_l)\right)\ ,
\ee
where we have assumed that the long mode is peaked at one frequency $k_l$.

Exactly the same expression can be computed in standard perturbation theory~\cite{Bernardeau:2001qr} as the
coupling between an general short and a spherically symmetric long mode
$\delta(k_l)$ leading to
\begin{align}
\delta^{(2)}_s(\vec k_s)=&\delta^{(1)}_s(\vec k_s)+\int\frac{\derd
\Omega_l}{4\pi} F_2(\vec k_s,\vec
k_l)\delta^{(1)}_s(\vec
k_s)\delta^{(1)}_l(k_l)=\delta^{(1)}_s(\vec
k_s)\left(1+\frac{34}{21}\delta^{(1)}_l(k_l)\right)\ .
\end{align}
We see that for long modes sufficiently far within the horizon so that a
Newtonian treatment is possible, the two expressions agree.

\section{The Coordinate Frame of $N$-body Simulations}

Usually, $N$-body simulations are performed on very small scales compared to the Hubble scale, and no hint is usually given onto in what gauge the calculation is actually performed. Further, the equations that are solved in the simulations are not even the General Relativistic equations, but the Newton's equations, where all the General Relativistic effects are neglected~\footnote{The fact that usually the spatial coordinates are rescaled by a time-dependent factor equal to the scale factor, using the so-called comoving coordinates, should not be misleading: that is just a convenient  change of variables for the same equations, which are still just the Newtonian ones.}.

Of course, there is a good reason for this. Usually simulations are performed in
boxes which are much smaller than the Hubble scale. Since all General
Relativistic effects, from the corrections to Newton's equation to the
specification of the coordinate frame, scale proportionally to $(H a)/k$, these
effects are usually negligible. We begin to need to worry when the box size of
the simulations becomes larger and larger, and reaches the Hubble scale. At this
point, at least naively, we have to modify our codes to include the General
Relativistic equations, choose some gauge in which to perform the calculation,
take care of what is actually the observable quantity that needs to be computed.
This is in fact different from $\delta\rho_m/\rho_m$, $\rho_m$ being the matter
density, as recently stressed in \cite{Yoo:2009au}, due to lensing and redshift
distortion effects. But doing all of this may seem a bit too much: at the end of
day, we know that large scales evolve linearly, and it is only scales much
smaller than the horizon that become non-linear and require $N$-body
simulations. Further, if the sound horizon is much smaller than the Hubble
scale, local {\it dynamics} does not really probe long distances, but it only
probes distances of the order of the mean free path of the particles, which is
the non-linear scale, and so it should not be affected by General Relativistic
effects. On small scales, we should be able to apply the Newtonian
approximation, and so our way of doing simulations should be fine to describe
the small scale non-linearities. There seems to be a tension between including
long wavelength fluctuations in the simulations, and the fact that the
non-linearities occur just on small scales.

{This tension has been solved in a recent paper~\cite{Chisari:2011iq}, where it was shown how, exploiting the above facts, it is possible to re-interpret the results of current Newtonian $N$-body simulations directly in the General Relativistic context, by providing a mapping between the results of $N$-body simulations and the fluctuations in a specific gauge valid at arbitrary length scales. The only mistakes in this procedure are suppressed by powers of $(v/c)\ll 1$, with no corrections of the form $(H a)/k$.

We are now going to argue that this same tension between $N$-body simulations and General Relativistic effects can be resolved in yet another way, by simply stating that in order to include long wavelength modes into the simulations, it is not necessary to make large-box simulations, but it is simply necessary to perform small-box simulations, in slightly curved backgrounds. The results obtained from the small scale simulations can then be reinterpreted as results obtained in local patches of the whole Universe. We will provide such a mapping~\footnote{The statement that in order to include large scale modes into small-box simulations one should include curvature and  a rescaling of the coordinates has been already given in~\cite{Tormen:1995sd} and then more properly in~\cite{Cole:1996hb}. However, a mapping from the frame of the simulations to the global frame had not been given, nor, it seems to us, a clear derivation has been presented. Further, all the statements in \cite{Tormen:1995sd,Cole:1996hb} are not in the General Relativistic context. All of this becomes important if we are dealing with modes comparable to Hubble size.}.

Let us see how this works by showing that simulations can be interpreted in the General Relativistic context as nothing but solving the Einstein equations in the frame defined by the local coordinates (\ref{eq:LocalGlobalMapping}), where the metric has the form (\ref{eq:FermiFRW}) with the scale factor, the Hubble rate and the curvature as given by (\ref{eq:localvariables}). If we now add short scale perturbations $\delta\Phi$ to the metric we have:
\bea
ds^2&=& -\left[1-\left(\dot H_L(t_L)+H_L(t_L)^2\right)r_L^2+2\delta\Phi(\vec
x_L,t_L)\right]dt_L^2\\ \nonumber
&&+\left[1-\frac{1}{2}\left(H_L(t_L)^2+\frac{K}{a_L(t_L)^2}
\right)r_L^2-2\delta\Phi(\vec x_L,t_L)\right]d\vec x_L^2\ .
\eea
As we write down the Einstein equations for a Universe of dark matter particles plus a cosmological constant, where the perturbations are non-relativistic, we immediately realize that in the above metric the Newtonian approximation is valid: the metric looks like Minkowski with just small corrections, and the system is non-relativistic. Straightforward algebra then shows that the Einstein equations take the form of the simple Poisson equation
\be\label{eq:simulation1}
\nabla^2\delta\Phi(\vec x_L,t_L)=4\pi G\delta\rho(\vec x_L, t_L)\ ,
\ee
while the geodesic equation for the dark matter particles takes the form
\be\label{eq:simulation2}
\ddot{\delta \vec x}(t_L)+2H_L(t_L)\dot{\delta \vec x}(t_L)=-\vec{\nabla}\delta\Phi(\vec x(t_L),t_L)\ .
\ee
In obtaining the above two equations, we have done several approximations and definitions that require explanation. We have defined 
\be
\delta\rho(\vec x_L, t_L)=\rho(\vec x_L, t_L)-\frac{3}{8\pi G} \left(H_L(t_L)^2+\frac{K}{a_L(t_L)^2}-\frac{\Lambda}{3}\right)\ , \qquad \dot{\vec x}(t_L)=H_L(t_L)\vec x+\dot{\delta \vec x}(t_L)\ ,
\ee 
where here we decided to focus on a $\Lambda$CDM Universe, thought we stress
that trivial generalization of our formulas apply to the case of clustering dark
energy~\cite{Creminelli:2006xe}.
Notice that the unperturbed velocity is nothing but the Hubble flow as seen at
small distances from the origin. Then, we have expanded in perturbations by
applying the Newtonian approximation: i.e., we have counted the
perturbations in powers of $\delta\Phi\sim v^2$, where $\vec v=\dot{\vec
x}(t_L)$, and taken the linear equations in these perturbations. Notice that
this amounts to taking the leading terms also in $r_L^2$ in the Einstein
equations, while we have not expanded in $\delta\rho/\rho$. The fact that these
approximations are justified can be checked a-posteriori,  but will become clear
in the next paragraph.

In fact, eqs.~\eqref{eq:simulation1} and \eqref{eq:simulation2} are {\it
exactly the same equations} that are solved in $N$-body numerical simulations.
This tells us two important things. First, that the Newtonian approximation is
indeed justified. Second, most importantly, we {\it now know} how to interpret
the above equations in a General Relativistic setting: they are the equations
for a local patch described by the local frame. Thanks to the change of
coordinates in (\ref{eq:LocalGlobalMapping}), we can interpret the results of
the $N$-body simulations as  points in the full manifold of the spacetime (let
us say for example as described in standard Newtonian gauge).

The presence of a long-wavelength mode affects the result of the $N$-body simulations in two different ways: first it affects the mapping from the global to the local coordinates in (\ref{eq:LocalGlobalMapping}), second it affects the evolution of the short modes by adding a small curvature (\ref{eq:localvariables}) to the effective local FRW Universe.

In summary, what we found can be synthesized by stating the following simple
procedure for performing $N$-body simulations that include large scale
fluctuations. Simulations are to be thought of as computing the gravitational
structures in the local frame defined by the change of coordinates
(\ref{eq:LocalGlobalMapping}). In the presence of a long-wavelength mode,
simulations should be performed in a curved (background) Universe where the
curvature is given by (\ref{eq:localvariables})~\footnote{As we stressed, the
same approach can be generalized to include perturbations at non-linear level
and to compute non-scalar quantities: in this case the local patch will not
evolve as a curved FRW.}. Any scalar quantity measured in the simulations, let
us say the proper number density of halos of a given mass, should be intepreted
as given at this time:
\be
N{\rm-body\ Simulations} \qquad \rightarrow\qquad n_p^L(\vec x_L,
t_L;\Omega_K(\zeta))\ ,
\ee
where the explicit dependence on $\zeta$ comes from the curvature, and the superscript $^L$ reminds us that the output of the $N$-body simulations is to be interpreted as given in Local coordinates.  From the mapping (\ref{eq:LocalGlobalMapping}), we then finally get the value in the set of coordinates that are globally valid, for example in Newtonian gauge:
\be\label{eq:passage of coordinates}
n_p^G(\vec x_G, t_G;\zeta)=n_p^L\left(\vec x_L(\vec x_G,t_G),t_L(\vec
x_G,t_G);\Omega_K(\zeta)\right)\ ,
\ee
where the superscript $^G$ reminds us that this quantity is defined in global coordinates valid everywhere, and  we have used that the proper number density is a scalar. 

Finally, we should comment on the initial conditions for the patches corresponding to the regions of space simulated in the $N$-body simulations. In the case of Gaussian initial conditions (we will comment on non-Gaussian initial conditions in the next section), it will turn out that to a very good approximation the initial power spectrum, expressed in terms of the local coordinates, should be the same as it would be in the absence of the long-wavelength mode. In order to understand the reason of this, it is useful to express the global metric in the comoving ($\zeta$) 
gauge which is comoving with the density perturbations (see appendix \ref{app:symbols}). In this gauge, for adiabatic initial conditions, and for modes that are far outside of the sound horizon, the metric takes the form
\be
ds^2=-dt^2+a^2 e^{2\zeta}d\vec x^2\ .
\ee
Let us decompose the fluctuation $\zeta$ in a long-wavelength and a
short-wavelength component $\zeta_l+\zeta_s$, where $_l$ stays for long, and
$_s$ stays for short. Let us assume for the moment that the long-component is on
scales longer than the sound-horizon. This means that it entered the Hubble
scale after matter-radiation equality. In this case, $\zeta_l$ is constant in
time. The property of the exponential is such that ${\rm Exp}(\zeta)={\rm
Exp}(\zeta_l)+{\rm Exp}(\zeta_s)$, which implies that in the limit in which we
can neglect completely the gradients of $\zeta_l$, $\zeta_l$ can be re-absorbed
in a constant rescaling of the scale factor, and is therefore unobservable. This
implies that the local physics (from matter radiation equality to recombination
and so on) happens in exactly the same way as if the long mode was absent. As we
learned in the former section, when we consider gradients of $\zeta_l$, the
leading effect of the long mode is to induce a curvature for the local Universe,
which clearly affects the local evolution. So, the initial power spectrum of the
short scales modes is the one that is obtained in a curved FRW Universe where
the curvature is given by the Laplacian of $\zeta_l$ as in
eq.~(\ref{eq:localvariables}). In practice, this means that we should run
numerical codes as CMBFAST~\cite{Seljak:1996is} or CAMB~\cite{Lewis:1999bs}, run
them with the relevant curvature of the Universe, and, after a rescaling by the
scale factor, simply interpret the output as in local coordinates. In reality,
it is not even necessary to obtain the power spectrum in such a curved Universe
as it is easy to realize that  the initial curvature is negligible. The
relevance of the curvature scales as
$\nabla^2_G\zeta_l/(a_G^2H_L^2)\propto1/\dot a_L^2$, and therefore it becomes
irrelevant in the past. In practice, neglecting the effect of the initial
curvature amounts to neglecting terms of order $\nabla^2_G\zeta_l(t_{L,\, in})$,
where $t_{L,\, in}$  is the initial time of the $N$-body simulation. When we
later define the bias we will define it as the coefficient of proportionality
between the local number density and $\nabla^2_G\zeta_l(t_{L,\, obs})$, where
$t_{L, obs}$  is the time of observation. The effect of the initial term scales
as $\dot a_L(t_{L,\,obs})^2/\dot a_L(t_{L,\, in})^2$ and gives a negligible
contribution to the bias if the initial time of the $N$-body simulation is early
enough. In practice, this is the simple fact that the curvature is irrelevant at
early times. This implies that, for long modes that entered the horizon during
matter domination, the initial conditions for the simulations are equivalent to
the ones in an unperturbed Universe.

The situation becomes slightly more complicated for long wavelength modes that
enter the horizon during radiation domination. In this case, there is a window
of time from horizon re-entry to matter-radiation equality during which
$\zeta_l$ depends on time. This means that the mode in this case can not simply
be interpreted as a rescaling of $a$ and an additional curvature term. In this
case gradients of the long fluctuation are relevant, as the mode travels
approximately an Hubble horizon in an Hubble time. In order to evaluate the
effect of the long mode on the short scale power, one should then solve the
non-linear equations that couple $\zeta_s$ and $\zeta_l$, along the line of what
done in \cite{Fitzpatrick:2009ci}. However, we can argue that this effect is
negligible. The biasing of structures as due to a long wavelength mode is an
intrinsically non-linear effect, and it therefore receives most of its
contribution from late times, as density perturbations become closer and closer
to being non-linear. In perturbation theory, it is straightforward to realize
that neglecting the non-Gaussianities of the initial conditions set up at a time
parameterized by $a_{in}$ amounts to neglecting a non-Guassianity of the matter
fields at a late time parametrized by $a_{obs}$ that is of the order of
$a_{in}/a_{obs}$. This is equivalent to  the order of the relative error in the
bias we have if we neglect the non-Gaussianity in the initial conditions. By
taking the initial conditions to be early enough, we can make this error small
enough. Given the fact that it is quite hard to measure the bias to great
precision, the initial condition can be set up at a reasonably late time.

Let us summarize the discussion about the initial conditions. Concerning modes that entered the horizon during matter domination, one can simply take the power spectrum in local coordinates as in an unperturbed FRW Universe. Concerning modes that entered the horizon during radiation domination, one should take non-Gaussian initial conditions that can be estimated in perturbation theory as for example in~\cite{Fitzpatrick:2009ci}; however, their effect is likely to be negligible.  The procedure we have outlined in this section enables to extract information about very long wavelength modes without practically modifying the $N$-body codes, and without having to run very large and time-consuming simulations. This should 
give a valid description for certain questions, such as the halo mass function, where spherical symmetry that we assumed to derive
Fermi coordinates is likely to be valid (see app.~\ref{app:Fermi-wave} for a
plane wave case).
In app.~\ref{app:local_expansion} and~\ref{app:local_parameters} we give a
detailed recipe for how to run a $N$-body simulation given the
cosmological parameters and the amplitude of the long-wavelength mode.

\section{Bias in General Relativity and its Scale Dependence}

As an application of our technique we will derive an expression for the bias
that is valid in the General Relativistic setting. As it has been recently
noted in~\cite{Dalal:2007cu,Slosar:2008hx} in the case of the local kind of
non-Gaussianities parametrized by the parameter $f_{\rm NL}^{\rm loc.}$, the
bias on large scales (as usually measured with respect to to the local matter
overdensity) receives a contribution that is scale dependent, proportional to
$1/k^2$, where $k$ is the wavenumber of the long-wavelength mode, proportional
to  $f_{\rm NL}^{\rm loc.}$. The same is expected to be true for the new non-Gaussian shapes that have been found in the Effective Theory of Multifield Inflation~\cite{Senatore:2010wk} (a generalization of the Effective Field Theory of inflation~\cite{Cheung:2007st}) that have support both in the equilateral and in the squeezed limit. These results were derived in the Newtonian
approximation, and here we will derive their generalization for wavelengths
comparable or longer than the horizon.

\subsection{Gaussian Bias}
If we consider surveys that are comparable to the horizon scale, then
relativistic effects become important and one needs to be very careful in
defining observables. We do not directly observe the proper number
of galaxies at a given point $n_p(t_G,\vec x_G)$ because the photons are
deflected and redshifted on their way from the source galaxy to the observer.

What we can do, is count the number of galaxies in bins of angle and redshift.
We will refer to the observed number density of galaxies, i.e., the number of
galaxies divided by the observed volume, as $n_{obs}(z,\theta,\phi)$.
Here $z$ is the observed redshift of the bin, and the tuple ($\theta,\phi$)
represents the observed angular position.\\
The observed position $(z,\theta,\phi)$
corresponds to a set of global coordinates $(t_G,\vec x_G)$.
Here we make use of the fact that a spacetime point can be described in in
different coordinate systems and that global, local and observed coordinates
are just three choices of such a coordinate frame that describe the same point.
Thus the global coordinates are a function of the observed coordinates
\be
(t_G,\vec x_G)=\left(t_G(z,\theta,\phi),\vec x_G(z,\theta,\phi)\right)\ ,
\ee
and since the proper number density $n_p$ is a scalar, i.e., a function of the
point rather than its coordinates, we have
\be
n_{p}(z,\theta,\phi)=n_p\bigl(t_G(z,\theta,\phi),\vec
x_G(z,\theta,\phi)\bigr)\ .
\ee
To compute the observed number density $n_{obs}(z,\theta,\phi)$ we need to
model both the proper density of objects $n_p$ and the mapping between proper
and observed coordinates. Let us start with the proper number density.
\subsubsection{Proper and Observed Number Density}
We have argued that in presence of long wavelength modes, the local
inertial frame corresponds to a homogeneous curved FRW Universe. As a result the
proper number density of galaxies at the spacetime point is given by the number
density in the effective curved Universe. We will denote this number $n_p(t_L;
\Omega_K)$. The time argument $t_L$
stresses the fact that the proper time of the free falling observer is in
general different from the global coordinate time.

We have:
\be\label{nphom}
n_{p}(z,\theta,\phi)=n_p\bigl(t_L\bigl(t_G(z,\theta,\phi),\vec
x_G(z,\theta,\phi)\bigr); \Omega_K\bigr)\ ,
\ee
where $t_L(t_G,\vec x_G)$ denotes the time in the Fermi frame centered at
$(t_G,\vec x_G)$ and $\Omega_K$ is the curvature associated with the long
wavelength mode. To evaluate this expression we need to compute the relation
between $(z,\theta,\phi)$ and $t_L$. We can split this relation in two parts.

First we can relate  $(z,\theta,\phi)$ to the global coordinates. As shown in
\cite{Yoo:2009au} there is a lapse between the coordinate redshift
$1+z_G=1/a_G(t_G)$ and the observed redshift $z$
\be
z-z_G=(1+z_G)\delta z_{G\to z}
\ee
where $\delta z_{G\to z}$ is given in
app.~\ref{app:symbols}. We also need to relate the global time coordinate to the
time in the Fermi frame at the origin (see eq. \ref{eq:LocalGlobalMapping}),
\be
t_L(t_G,\vec x_G)=t_G+\delta t_{G\rightarrow L}(t_G,\vec x_G). 
\ee
The time shift between the global and the local coordinates is the difference
between the global coordinate-time and the proper-time. In Newtonian gauge we
have
\begin{align}
\delta t_{G\to
L}(z,\theta,\phi)=&t_L\bigl(t_G(z,\theta,\phi)\bigr)-t_G(z,\theta,\phi)=\int_0^{
t_G(z,\theta, \phi) } \Phi(t_G'(z,\theta,\phi),\vec x_G(z,\theta,\phi))dt_G'\ ,\\ \nonumber
=&-\frac{1}{H(z)}\left[\zeta\bigl(t_G(z,\theta,\phi),\vec
x_G(z,\theta,\phi)\bigr)+\Phi\bigl(t_G(z,\theta,\phi),\vec
x_G(z,\theta,\phi)\bigr)\right]\ .
\end{align}
We can now expand eq.~\eqref{nphom} to first order in the perturbations to
obtain,
\begin{align}
\label{eq:3}
n_p(t_L;\Omega_K)=& n_p(t_L, \Omega_K=0)\left[1+\frac{1}{\bar
n_{p}}\frac{\partial
n_p}{\partial \Omega_K} \Omega_K \right]\ ,
\end{align}
where $\bar n_{p}$ is the unperturbed number density at the redshift of
observation.
Doing so, we have performed a split into background and perturbation
such that $n_p(t_L, \Omega_K=0)$ is not a scalar but a function of its time
argument. Thus
\begin{align}\label{eq:3}
n_p(z,\theta,\phi)=& n_p(t_G, \Omega_K=0)\left[1+\frac{1}{\bar
n_{p}}\frac{\partial n_p}{\partial \Omega_K} \Omega_K+\frac{\partial \log\bar
n_p}{\partial
t}\delta t_{G\rightarrow L}\right]\\
=&n_p(z_G, \Omega_K=0)\left[1+\frac{1}{\bar n_{p}}\frac{\partial
n_p}{\partial \Omega_K} \Omega_K+\frac{\partial \log\bar n_p}{\partial
\log (1+z)}\delta z_{G\rightarrow L}\right]\ ,
\end{align}
where we have rewritten the prefactor and the time shift in terms of the
global redshift $z_G$, which is possible
since there is a one-to-one relationship between redshift and time in the
auxiliary background Universe that can be translated into a relation between
$\delta t_{G\to L}$ and $\delta z_{G\to L}$
\be
\delta t_{G\to L}(z,\theta,\phi)
=-\frac{z_G(t_L)-z_G(t_G)}{H(z)(1+z)}
=-\frac{\delta z_{G\to L}}{H(z)}\ .
\ee
When calculating spherical averages, the observed redshift $z$ is
fixed while coordinate redshift $z_G$ and global time $t_G$ vary. As we will see
shortly, it is beneficial to evaluate the prefactor at $z=z_G+\delta z_{G\to
z}$
\be
n_p(z,\theta,\phi)=n_p(z, \Omega_K=0)\left[1+\frac{1}{\bar n_{p}}\frac{\partial
n_p}{\partial \Omega_K} \Omega_K+\frac{\partial \log\bar n_p}{\partial
\ln (1+z)}\left(\delta z_{G\rightarrow L}-\delta z_{G\rightarrow
z}\right)\right]\ .
\ee
We can now define the bias as
\be\label{eq:bias_def}
b_{\Omega_K}(t)=- \frac{1}{\bar n_{p}}\frac{\partial
n_p}{\partial \Omega_K},
\ee
and use that $\Omega_K(t)=2\nabla^2_G\zeta/(3a^2 H^2)$ in eq.~\eqref{eq:3}. We
discuss the relation between this definition of the bias and the standard one in the Newtonian approximation in
the next section.
\subsubsection{Volume Distortion}
Finally, to compute $n_{obs}(z,\theta,\phi)$ we need to take into account
the distortions in the volume induced by the mapping between $(z,\theta,\phi)$
and the local frame. These geometric factors were recently derived at linear
level in~\cite{Yoo:2009au}. We denote $V_p$ the proper volume corresponding to a
bin in $(z,\theta,\phi)$ and define
\be
V_p=\bar V_p (1+{\cal J}),
\ee
where $\bar V_p$ is the corresponding volume in an unperturbed Universe
and\footnote{Our expression for ${\cal J}$ assumes that the survey is volume
limited. If instead the survey is flux limited, we have to add the
corrections due to the change in the apparent luminosity. In this case we have
to replace ${\cal J}$ with
\be
{\cal J}\quad\rightarrow\quad {\cal J}-5p\,\delta{\cal D}_L\ .
\ee
See app.~\ref{app:symbols} for details.
} 
\be\label{eq:projection}
{\cal J}=-\Phi-(1+z)\frac{d}{d z}\delta z_{G\rightarrow z}-2\frac{1+z}{H r}\delta z_{G\rightarrow z}-\delta z_{G\rightarrow z}-2
\kappa+\frac{1+z}{H}\frac{d H}{dz}\delta z_{G\rightarrow z}+2 \frac{\delta r}{r}\ ,
\ee
gives the geometrical projection effects computed in \cite{Yoo:2009au}.
Finally, we have
\be\label{eq:final}
n_{obs}(z,\theta,\phi)=n_p(z, \Omega_K=0)\left[1-b_{\Omega_K}
\Omega_K+\frac{\partial \log\bar n_p}{\partial
\ln (1+z)}\left(\delta z_{G\rightarrow L}-\delta z_{G\rightarrow
z}\right)+\mathcal{J}\right]\ .
\ee
Note that all the terms in the bracket are first order, i.e., they can be
evaluated at $z$, $z_G$ or $t_G$ equivalently, since these agree at zeroth
order.
\subsubsection{Observed Overdensity \& Averaging}
The observed overdensity is the fractional difference between the overdensity
in a certain direction and the angular average over the survey area
\be
\delta_{obs}(z,\theta,\phi)=\frac{n_{obs}(z,\theta,\phi)-\bar{n}_{obs}(z)}{\bar
{ n}_{obs}(z)}\ .
\ee
When evaluating the observed mean number density we can use that all the
terms in the bracket in eq.~\eqref{eq:final} vanish, when averaged over a
sufficiently big survey area. Hence we obtain for the angular average
\be
\bar{n}_{obs}(z)=\int_{\Omega_{survey}} \frac{\sin{\theta}\derd \theta \derd
\phi}{\Omega_{survey}}
n_{obs}(z,\theta,\phi)=n_{p}(z;\Omega_K=0)\ .
\ee
Now, the benefit of evaluating prefactor in eq.~\eqref{eq:final} at the
observed redshift becomes obvious. Since the observed redshift is fixed,
$n_{p}(z,\Omega_K=0)$ agrees with the survey average and we have for the
observed overdensity (we will ignore the additional effects on monopole and dipole, which 
are influenced by the contributions at the observer's position),
\be\label{eq:observation}
\delta_{obs}(z,\theta,\phi)=-b_{\Omega_K} \Omega_K+\frac{\partial
\log\bar n_p}{\partial
\ln (1+z)}\left(\delta z_{G\rightarrow L}-\delta z_{G\rightarrow
z}\right)+\mathcal{J}\ .
\ee
The volume distortion is in principle observable and
thus has to be gauge invariant by itself. The first term
$-b_{\Omega_K}\Omega_K$ is the number of collapsed objects in the inertial
frame and thus totally independent of the choice of coordinates on the global
manifold. The remaining redshift lapse is gauge invariant as we show in
app.~\ref{app:symbols}. Together with the first term it forms another
observable.
With the above results the observed number density can be written as
\be
n_{obs}(z,\theta,\phi)=\bar
n_{obs}(z)\left[1+\delta_{obs}(z,\theta,\phi)\right]\ .
\ee
We can also relate the expression in (\ref{eq:final}) to the overdensity in the
global coordinates,
$\delta_G(t_G,\vec{x}_G)=(n_p(t_G,\vec{x}_G)-\bar{n}_p(t_G))/\bar{n}_p(t_G)$,
where the averaging is done over hypersurfaces of constant coordinate time. We
obtain
\be
n_{obs}(z,\theta,\phi)= \bar n_{obs}(z)\left[1+ \delta_G -
\frac{\partial \log\bar n_p}{\partial \log (1+z)}  \delta z_{G \to z} +
{\cal J}\right].
\ee
Note that in the case where the tracer has a number density that scales like
$(1+z)^{3}$ the combination $ \delta_G - ({\partial \log\bar
n_p}/{\partial \log (1+z)})  \ \delta z_{G\to z}$ becomes  $\delta_G -
3
\delta z_{G\to z	}$ in agreement with \cite{Yoo:2009au}.

In our formalism it was natural to define the bias directly in terms of the
Laplacian of the $\zeta$ perturbation at the point of interest, which in turn is
proportional to the curvature of the local FRW Universe. Because of the
Friedmann equation, the curvature turns out to be proportional to the
overdensity of the Universe at the source galaxy position, as shown next. 
This offers us a procedure to extract the bias from $N$-body simulations: run
simulations with varying $\Omega_K$, and then take the derivative with respect
to this parameter.

\subsection{Comparison with Standard Newtonian Treatment of Bias}
Our bias definition tells us that we should take the derivative of
the
number density with respect to the curvature of the local Universe. While our
receipe is well defined in the full General Relativistic setup, it still should
agree in the limit in which the long mode is well inside the horizon, so that
the Newtonian approximation is valid for the long mode itself. However, in this
case a naive look at the expression might make us think that the two procedures
do not agree. Indeed, in the classical Newtonian treatment, the bias is defined
as the derivative of the number density with respect to the local
long-wavelength overdensity. In this section we will first relate the above bias
definition to an overdensity and then consider the subhorizon limit.

The curvature energy density of the local Universe scales as
$\Omega_K=\Omega_{K,0}H_0^2/(a_G H_G)^2$ and is thus fully specified by its
value at
redshift 0. The latter can be related to the matter density in (synchronous) comoving gauge
as
\begin{align}
\Omega_{K,0}=&-\frac{K}{H_0^2}=\frac{2}{3}\frac{\nabla^2
\zeta}{H_0^2}=-\left(1-\frac{f_0 H_0^2}{\dot{H}_0}\right)(1-\Omega_{DE,0}
)\delta_{l,0}^{(com)}\\
=&-\left(\Omega_{m,0}+\frac{2}{3}f_0\right)\delta_{l,0}^{(com)}
\end{align}
where the last two equalities are valid for a Universe with time varying dark
energy and a $\Lambda$CDM Universe, respectively.
Hence the bias term in eq.~\eqref{eq:final} can be written as
\begin{align}
-b_{\Omega_K}(t)\Omega_K(t)=&b_{\Omega_K}(t)\left(1-\frac{f_0
H_0^2}{\dot{H}_0}
\right)\frac{(1-\Omega_{DE,0})H_0^2}{D(t)H(t)^2a(t)^2}\delta_{l}^{(com)}(t)
\equiv b(t)\delta_{l}^{(com)}(t)
\end{align}
We restored the time dependence of the long wavelength density
perturbation, dividing by the linear growth factor $D(t)$.
 Our new bias
$b_{\Omega_K}$ is related to the standard bias parameter by a time dependent but
scale independent factor. From the above equation we can see that the
density perturbation in the comoving gauge is equally suited, at an algebraical
level, as an
expansion parameter for the galaxy bias, but the justification of this
statement relies simply on the proportionality of $b$ to $b_{\Omega_K}$.
Further, the bias expressed in terms of $\Omega_K$ makes manifest its
gauge-invariant physical origin and the fact that the biasing vanishes for modes
longer than the Hubble scale.

Well inside the horizon ($k\gg aH$) the velocity term in the relation between
comoving and Newtonian gauge matter overdensity (see eq.~\eqref{eq:densnewcom}
in app.~\ref{app:symbols}) becomes negligible and thus
both density perturbations reduce to the Newtonian density perturbation
$\delta_l^{N}\approx\delta_l^{com}\approx\delta_l$.
Furthermore, inside the horizon the volume distortion as well as the lapse
between the global, local and observed redshift are negligible. Thus
eq.~\eqref{eq:observation} reduces to
\begin{align}
\delta_{obs}(z,\theta,\phi)=&-b_{\Omega_K}(t) \Omega_K(t)=b(t)
\delta_l(t)\ ,
\end{align}
which is the standard relation between observed tracer overdensity and
underlying matter overdensity in the Newtonian approximation.

Finally, we point out that another way to understand the connection between our
bias $b_{\Omega_K}$ and the standard one is by referring to the peak background
split method. There, in the Newtonian context, it is usually assumed that the
presence of a long scale mode can be interpreted as a shift of $\delta_c$:
$\delta_c\ \rightarrow\ \delta_c-\delta_{l}$, and after Taylor expansion we
obtain the expression for the linear bias. In our context, the presence of a
long mode is instead interpreted as a curvature of the background Universe, and
therefore we have to rescale $\delta_c$ accordingly to $\delta_c(\Omega_K=0)\
\rightarrow\ \delta_c(\Omega_K\neq 0)$ and then Taylor expand. In
app.~\ref{app:curved_spherical_collapse}, we show that indeed the two approaches
are equivalent on short scales.

\subsection{Bias in Presence of non-Gaussianities of the Local Kind}
So far we have assumed that the only way a long-wavelength mode affects the
local structure formation is through its dynamical effects: that
is by changing the local geometry and by introducing curvature in the
resulting local FRW Universe.  If the initial conditions are Gaussian, this
accounts for all the effects of the long mode on local processes: in the
linear regime the statistical properties of the short wavelength modes are
decoupled
from long wavelength modes, and the non-linearities kick in only at late
times on small scales, where all the effect of the long mode can be absorbed
by a redefinition of the local expansion history. If the initial conditions are
non-Gaussian, then the statistical properties of the initial short scale
fluctuations are in general affected by the presence of a long mode and this
has to be taken into account. The scales that become non-linear are very
small compared to the horizon, and the scales that we are interested in are much
larger than the non-linear scale. Thus, in order for the properties of the short
scale fluctuations to be affected by the long mode, the non-Gaussian initial
conditions need to be such that they correlate very long and very short modes.

In general the description of the statistical distribution of modes in the
initial conditions requires knowledge of all the moments of their distribution.
For special cases a limited set of parameters $\bf p$ is sufficient. For
instance, if the initial conditions are Gaussian, they are fully quantified by
their variance. If the parameters $\bf p$ depend on the long wavelength
amplitude, then the proper number density of objects has an additional
explicit dependence on the long wavelength amplitude. Thus we can generalize
eq.~\eqref{eq:passage of coordinates} to:
\be
n_p(\vec x_G, t_G;\zeta)=n_p\bigl(\vec x_L(\vec x_G,t_G),t_L(\vec
x_G,t_G);\Omega_K,{\bf p}(\zeta)\bigr)\ .
\ee
These parameters $\bf p$ represent all the relevant information needed to
describe the initial conditions on small scales. The abundance of
objects of a given mass $M$ is mainly sensitive to the amplitude of fluctuations
smoothed on a scale enclosing the mass, given in terms of the variance
$\sigma_M$. There is also a weak dependence on the slope of the power spectrum
at the scale $M$ and possibly on parameters describing deviations from a
Gaussian distribution of the small scale modes, e.g.\ skewness. For definiteness
we will consider only the dependence on $\sigma_M$.

The so-called local kind of non-Gaussianities~\cite{Gangui:1993tt} that can be
produced in multifield inflationary models~\cite{Lyth:2002my,Zaldarriaga:2003my}
or in the new bouncing cosmology~\cite{Creminelli:2007aq} provides an example
where $\sigma_M$ depends explicitly on the long wavelength amplitude $\zeta$~\footnote{The same is expected to be true for the new non-Gaussian shapes that have been found in the Effective Theory of Multifield Inflation that have support both in the equilateral and in the squeezed limit~\cite{Senatore:2010wk}.}. In
these models the initial conditions are such that the curvature perturbation is
a non-linear function (local-in-space) of an auxiliary Gaussian random variable
$\zeta_g$:
\be
\zeta(\vec x_G)=\zeta_g(\vec x_G)-\frac{3}{5}f_{\rm NL}^{\rm
loc.}\left(\zeta_g(\vec x_G)^2-\langle\zeta_g^2\rangle\right)\ .
\ee
If we decompose $\zeta$ into long and a short modes as we did before, we can
see that the short mode takes the form
\be
\zeta_s\simeq\left(1-\frac{6}{5}f_{\rm NL}^{\rm
loc.}\zeta_{g,l}\right)\zeta_{g,s}\ ,
\ee
where we have neglected a term in $\zeta_s^2$ which is irrelevant for our
discussion of the bias. From this equation we see that the variance of the short
scale power is modulated by the long mode. This implies that in the case of
local non-Gaussianities there is an additional source of bias. If we set up the
initial conditions for the simulation in the presence of non-Gaussianities of
the local kind, the resulting proper number density of halos $n_p$ will depend
on the long-mode not only through its explicit dependence on the curvature of
the local Universe, but also through the dependence on the initial power
spectrum of the modes~\footnote{Though we are now talking about non-Gaussian
effects, notice that we are consistently treating the long mode at linear
level.}. Eq.~(\ref{eq:final}) is generalized to
\bea \nonumber
n_{obs}(z,\theta,\phi)&\simeq &n_p(z; \Omega_K=0,{\bf \bar p})\times
\\\nonumber
&&\biggl[1-
b_{\Omega_K} \frac{2\nabla^2_G\zeta}{3a^2 H^2}+\frac{1}{\bar{n}_p}\frac{\d
n_p}{\d \sigma^2_M}\frac{\d \sigma^2_M}{\d \zeta} \zeta 
+\frac{\partial \log\bar
n_p}{\partial \log (1+z)}\left(\delta z_{G\to L} - \delta
z_{G\to z}\right)+ {\cal J}\biggr]\\
&=&\bar{n}_p(z)\biggl[1-
b_{\Omega_K} \Omega_K+b_\zeta \zeta
+\frac{\partial \log\bar
n_p}{\partial \log (1+z)}\left(\delta z_{G\to L} - \delta
z_{G\to z}\right)+ {\cal J}\biggr]\ ,
\eea
where $\d\sigma_M^2/\d\zeta=-12f_{\rm NL}^{\rm loc.}/5 $ is independent of $M$
and $\bf \bar{p}$ describes the initial conditions in absence of long
perturbations. We see that in presence of non-Gaussianities of the local
kind the bias receives an additional contribution proportional to $\zeta$, while
the standard Gaussian contribution is proportional to $\nabla^2 \zeta$. There is
a relative scale dependence proportional to $k^2$ between the two. But this
does not imply the very unphysical result that the bias blows up as
$k\rightarrow 0$. It is rather the fact that the bias for large scales should be
interpreted as a different bias: as the coefficient of proportionality between
the local number density and $\zeta$ and $\nabla^2 \zeta$~\footnote{Our
conclusions about the bias as due to local non-Gaussianities are in general
agreement with the ones of~\cite{Wands:2009ex}, though they differ in the way
the results are derived and in parts of their interpretation.  We stress that
our derivation does not crucially rely on the assumption of spherical symmetry.
It should allow for a straightforward generalization to the non-linear case where
spherical symmetry can not be used.}.\\
The final expression for the observed overdensity in presence of local
non-Gaussianities is thus
\be
\delta_{obs}(z,\theta,\phi)=-b_{\Omega_K}\Omega_K+b_\zeta \zeta
+\frac{\partial \log\bar n_p}{\partial \ln (1+z)}\left(\delta z_{G\rightarrow
L}-\delta z_{G\rightarrow
z}\right)+\mathcal{J}\ .\label{eq:finalnongauss}
\ee
The presence of $\Phi$ terms in the redshift lapse terms and the volume
distortion term mimicks $f_{\rm NL}^{loc.}$
of order unity. But this should not bias any measurement of non-Gaussianity
since the General Relativistic effects are calculable and can thus be removed
from the measurement.
\begin{figure}[t]
 \centering
\includegraphics[width=0.49\textwidth]{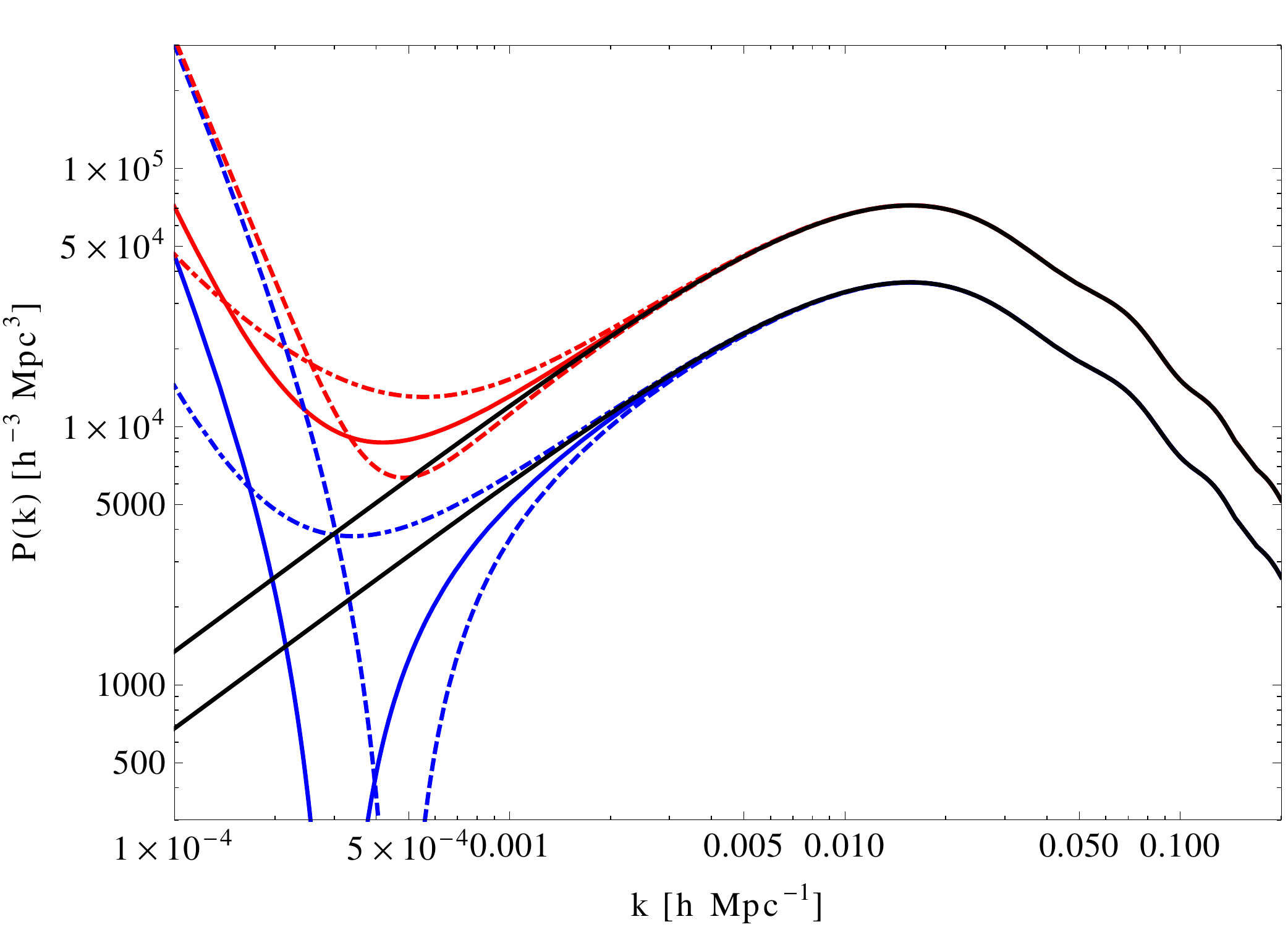}
\includegraphics[width=0.49\textwidth]{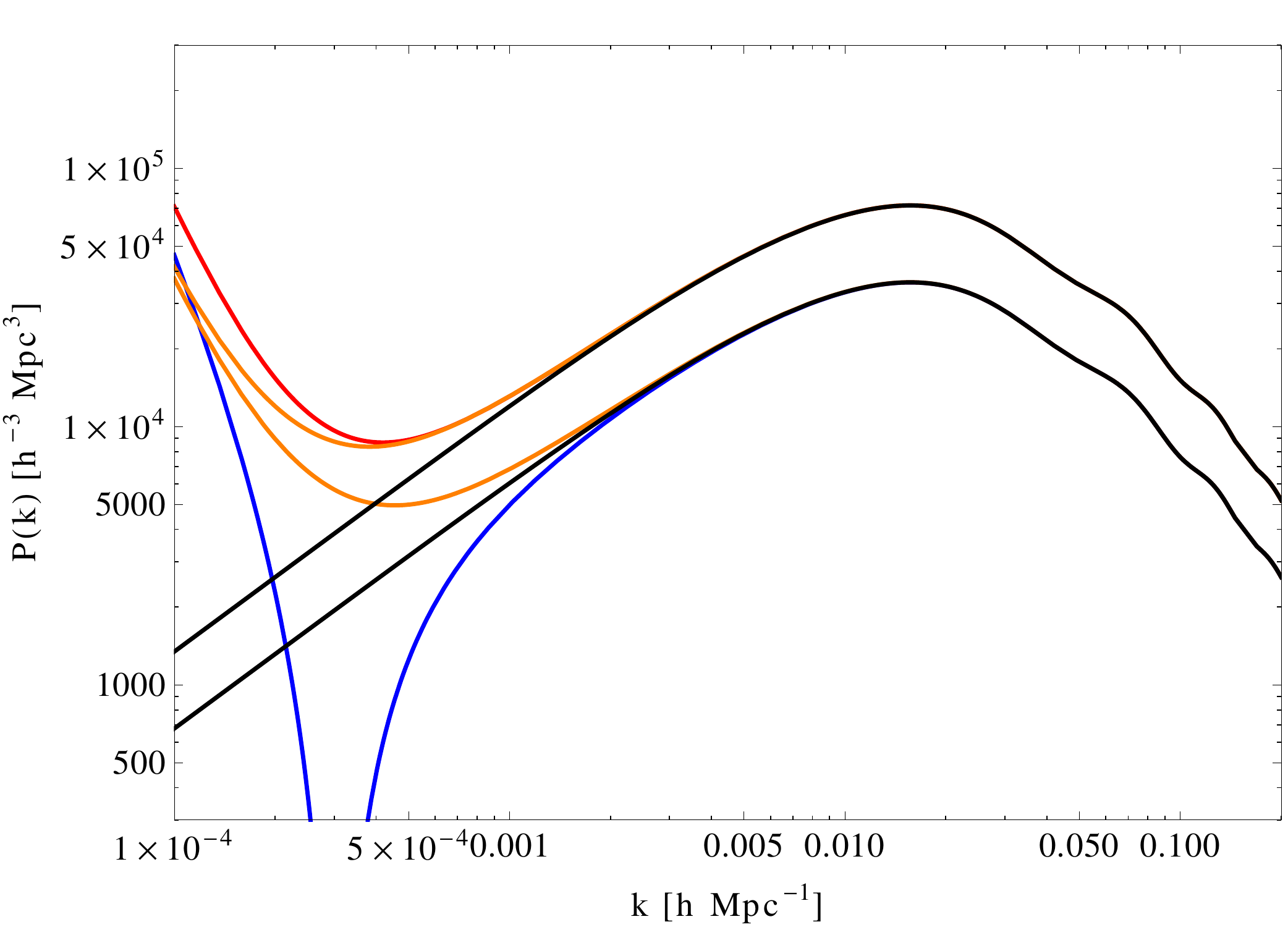}
\caption{Observed galaxy power spectrum for $z=1$, $b_{\Omega_K}=1.5$ ($b=2$)
and $\partial \log n_p/\partial \log (1+z)=3$. We choose the following cosmological parameters: $\Omega_m=0.28\, ,\;\sigma_8=0.84\, , \; H_0=0.70$. \emph{Left panel:} We show the
spectra parallel to the line of sight (red) and transverse to the line of sight
(blue). The solid line is for Gaussian initial conditions, whereas
dot-dashed is $f_{\rm NL}^{loc.}=+0.5$ and dashed is $f_{\rm NL}^{loc.}=-0.5$.
The lower black
line is just the power spectrum of density in comoving gauge, the upper is
multiplied by the redshift
space distortion factor $(1+f/b)^2$ to give the power parallel to the line of
sight.
We see that the effects of non-Gaussianity and GR-effects on the power spectrum differ, because the latter 
depend also on the line of sight parameter $\mu$ through the peculiar velocity effects.  \emph{Right panel:} Same as left, but orange lines show non -Gaussian power spectrum without the
GR-effects (just redshift space distortions).}
\label{fig:observedpower}
\end{figure}

As in the case of Gaussian initial conditions our formula can be applied
directly to the results of $N$-body simulations, but for illustrative purposes
we can also calculate the effect analytically assuming that the number density
of collapsed objects is described by a universal mass function
\be\label{eq:fitting}
n_{p}\propto f\left(\frac{\delta_c}{\sigma_M}\right)\ ,
\ee
i.e., it is a function of the peak height, the ratio of collapse threshold and
fluctuation amplitude.
In this case the derivatives of $n_p$ with respect to
$\sigma_M^2$ and with respect to the curvature $\nabla^2\zeta$ are related:
\begin{align}
\label{eq:gaussnongauss}
\frac{\d { n}_p}{\d \zeta}=& \frac{\d {n}_p}{\d
\sigma_M^2}\frac{\d \sigma_M^2}{\d\zeta}=-\frac{1}{2}\frac{\d {n}_p}{\d
\delta_c}\frac{\delta_c}{\sigma_M^2}\frac{\d \sigma_M^2}{\d\zeta}\ , \\
\nonumber
\frac{\d { n}_p}{\d (-\Omega_K)}=& \frac{\d { n}_p}{\d
\delta_c}\frac{\d\delta_c}{\d(-\Omega_K)}\ ,
\end{align}
which means that the Gaussian and the non-Gaussian bias are analytically related.

For general initial distributions, the additional contribution to the
fluctuations in the proper number density arise from $\partial {\bf
p}/\partial \zeta$. In the case of $\sigma_M$ considered above, this is nothing
but the squeezed limit of the three point function, because this measures
the coupling between small and large scale modes. Squeezed in this context
refers to the fact that we are talking about a correlation between short and
long wavelengths so two of the momenta in the relevant three point function are
very large compared to the other one and thus the three momenta form a squeezed
triangle. In the local model that we use as an illustration, the
derivative
\be
\frac{\d \sigma_M^2}{\d \zeta(k)}
\ee
is independent of both $k$ and of $M$ (we have explicitly pointed out that the
derivative might be different as a function of the wavenumber of the long
momenta). Relatively simple models can and have been constructed where this
derivative depends on both $M$ and/or $k$~\cite{Chen:2009zp}.  Even when this
derivative is not constant our formulas remain valid.

As an illustration, in fig.~\ref{fig:observedpower} we show an example for the
the observed
galaxy power spectrum for $b_{\Omega_K}\approx1.5$ ($b=2$) at $z=1$ assuming a
volume limited survey. The plots show the power parallel and orthogonal to the
line of sight for a sample with evolution slope of
$\partial \log \bar{n}_p/\partial \log (1+z)=3$. We are adding local
non-Gaussianity of $f_{\rm NL}^{loc.}=\pm1$. The right panel shows that ignoring
the
GR-effects could lead to a fake detection of $f_{\rm NL}=\mathcal{O}(1)$, but
this 
degeneracy is broken if modes both transverse and along the line of sight are considered. 
This is because GR-effects have a peculiar velocity contribution that has a $\mu$ dependence, 
where $\mu=\cos \theta$ and $\theta$ is the angle between the Fourier mode angle
and the line of sight.
For a related study on distinguishing GR effects from primordial non-Gaussianity
see \cite{Bruni2011}.
Note that
the magnitude of the GR-effects depends on the redshift distribution of the
sample. The non-Gaussian bias parameter $b_\zeta$ is calculated from the
Gaussian bias using eq.~\eqref{eq:gaussnongauss}. For the evaluation we are
neglecting all the line of sight integrals (convergence, Shapiro-delay,
integrated Sachs-Wolfe effect), which contribute power mainly to transverse
modes. For the details of the evaluation of the observed power spectrum, we
refer the reader to eq.~\eqref{eq:deltaobsk} in app.~\ref{app:symbols}.

It is also important to note that the relevant quantity is the change in the
amplitude of fluctuations at a given physical scale $M$ not of course a comoving
scale. In single field inflationary models this derivative goes to zero in the
squeezed limit, when $k$ corresponds to a much larger scale than $M$. In fact it
goes to zero as the square of $k$ just because the long wavelength mode affects
the production of the short modes during inflation only through tidal type
effects. In a sense it goes to zero in this way for reasons identical to the
ones that lead to the ${\nabla^2_G\zeta}$ dependence in the bias formulas.
Thus, in single field inflationary models there is no modulation of the proper
number density that scales with lower powers of $k$ than
${\nabla^2_G\zeta}$~\footnote{One can construct
examples were there is an intermediate ``squeezed regime" over which the scaling
is different than $k^2$ but for sufficiently large ratio the scaling needs to be
$k^2$~\cite{Silverstein:2008sg,McAllister:2008hb,Green:2009ds,Barnaby:2009mc,
Flauger:2009ab,Chen:2006xjb}.}.
The reader familiar with the standard calculation of the single-field inflationary three
point function might recall that in the squeezed limit they do not seem to
vanish but that they satisfy a consistency condition where the shape of the
three point function looks like that of a local model with an amplitude given by
the tilt of the fluctuations usually called $(n_s-1)$. But this dependence
arises entirely from the fact that what is being calculated is a three point
function in terms of comoving momenta. If expressed in terms of physical
momenta, the $(n_s-1)$ is exactly the amplitude required to make the
relevant derivative vanish.

\subsection{Observing Local-type non-Gaussianities in the Presence of GR Corrections}

The salient fact about the local-type non-Gaussianities is that they induce a
dependence of the proper number density of objects on the long wavelength modes
that is much stronger than what the dynamical effects can produce, proportional
to $\zeta$ rather than ${\nabla^2_G\zeta}$. Unfortunately when we count objects
in our Universe there are projection type effects that make the observed
densities depend directly on $\zeta$ even if the proper density does not.
The volume corresponding to a given observed range of angles and redshifts
varies as a result of the long wavelength modes and results in the factor of
$\cal J$ in eq.~\eqref{eq:final}. Furthermore a given observed redshift
corresponds to a different proper time in different directions resulting in the
terms proportional to ${\partial \log\bar n_p}/{\partial \log (1+z)}$ . Both of
these
terms lead to contributions proportional to $\zeta$, contributions that have the
same form as that coming from the local-type of non-Gaussianities. Failing to
correct for them would bias the results for $f_{\rm NL}^{\rm loc.}$ by a number
of order one which depends on the details of the population of objects surveyed.

Of course the various terms have different dependences on the properties of the
objects as they depend on different derivatives of  $n_p^L$. The effects will
also depend differently on redshift and furthermore, because the GR effects are
projection effects induced by  the intervening matter, it may be possible to
distinguish them using observations of the distribution of matter at the
intervening redshifts. It is beyond the scope of this paper to quantify the
extent to which these different effects may be isolated in practice or what it
is
required of the observations to distinguish them.

It is clear however that the GR effects are just projection effects, so if we
were able to construct observables that were directly sensitive to quantities in
the local frame we could side track those difficulties. In this section we just
want to point out that this is in principle possible. We will not address wether
this can be done in practice given our current tools or wether this route is
better than just trying to correct for the projection distortions in a realistic
situation.

To be able to ignore the projection effects we would need to be able to measure
the proper density of some object at a given proper time. Thus we would need a
ruler that would allow us to measure distances independently of the observed
angles and redshifts and we would need a clock that would allow us to compare
regions of the Universe at the same proper time independently of the observed
redshift. If we managed to find such local clocks and rulers the observed
density should only depend on ${\nabla^2_G\zeta}$ in the absence of primordial
non-Gaussianity. In fact there should only be a ${\nabla^2_G\zeta}$ dependence
in any single field model of inflation.

There are many such rulers that one could imagine using. One option is to use
the acoustic scale. This could be used for example by measuring the the number
of objects in regions of a given size in units of the acoustic scale. The
acoustic scale can be determined by measuring the correlation function of these
or other objects. Another option is to measure the ratio of the densities of two
tracers. Then the volume projection effects would cancel, in a sense we are
using the density of one of the objects to define the ruler for the other.

We still need a clock to make sure that one is comparing the number densities at
a fixed proper time rather than observed redshift. This difference is
responsible for the terms proportional to $({\partial \log\bar n_p}/{\partial
\log (1+z)})$ in eq.~\eqref{eq:final}. This appears a bit more tricky but
not a problem of principle. One needs to date the object observed independently
of their redshift, something that happens automatically for tracers that appear
only at a characteristic time in the history of the Universe. Examples of such
things might one day be the first stars or perhaps quasars could be used as
their abundance has a peak in redshift. In other words, the ratio of densities
of tracers that come from a given proper time and could be identified without
using the observed redshift would only depend of the long wavelength modes
through  the ${\nabla^2_G\zeta}$.

A similar construction for measuring the three point function in the squeezed
limit  could be accomplished using the CMB. The CMB comes already from a defined
proper time, the recombination of hydrogen provides the clock. So one could use
the dependence of the small scale power on large modes as a test of the squeezed
three point function. One should use a local definition for fluctuations and
power, meaning normalizing the fluctuations to the mean fluctuation level in the
region of interest to eliminate the equivalent of the $\delta z_{G\to z}$
 term in our equations for the densities of haloes. There is
still the projection effect related to the mapping between angles and physical
distances at recombination. This however can be avoided by comparing the
amplitude of fluctuations at a fixed scale measured in units of the acoustic
scale, thus at a fixed physical scale. For example the amplitude of the power
spectrum at the $N$-th peak should only depend on the long modes through the
${\nabla^2_G\zeta}$ in the absence of primordial non-Gaussianities. One could
also use the anisotropies in the small scale power to de-lense the CMB along the
lines considered in~\cite{Hirata:2002jy,Hirata:2003ka}.

\section*{Acknowledgments}
While this paper was being written, very recently
Ref.~\cite{Challinor2011,Bonvin2011,Jeong:2011as,Bruni2011} appeared
which treat problems similar to ours and reach similar conclusions where there
is overlap with our study.
We would also like to thank the Asian Pacific Centre for Theoretical
Physics in Pohang, Korea, for their kind hospitality during the workshop on
``Cosmology and Fundamental Physics". TB thanks the Lawrence Berkeley National
Laboratory, the Berkeley Center for Cosmological Physics, and EWHA Womans
University for kind hospitality while parts of this project have been carried
out.
We thank Jaiyul Yoo for helpful discussions.
This work is supported by DOE, the Swiss National Foundation under contract
200021-116696/1 and WCU grant R32-2009-000-10130-0.

\section*{Appendix}
\appendix

\section{Fermi Coordinates from $\zeta$-gauge}\label{app:Fermi-zeta}

Here we give the change of coordinates necessary to go from $\zeta$-gauge to the Fermi coordinates in the case of a spherically symmetric perturbation.

In $\zeta$-gauge the metric takes the form
\be
ds^2=-N^2dt^2+\delta_{ij}e^{2\zeta}a^2\left(dx^i+N^i dt\right)\left(dx^j+N^j dt\right)\ ,
\ee
where we have used the ADM parametrization. In this gauge time diffeomorphisms
are fixed by requiring~$T^0_i=0$.
The lapse $N$ and shift $N^i$ are constrained variables, whose solutions in terms of~$\zeta$ are~\cite{Seery:2005wm}
\be\label{eq:constraint}
N=1+\frac{\dot\zeta}{H}\ ,\qquad N_i=-\frac{\nabla_{G,i}\zeta}{H}-\frac{\dot H}{H^2}\frac{a^2}{c_s^2}\frac{\nabla_{G,i}}{\nabla_G^2}\dot\zeta\ .
\ee
The equation of motion for $\zeta$ reads~\cite{Seery:2005wm}
\be
\frac{1}{a^3}\d_t\left(\frac{a^3}{c_s^2}\frac{\dot H}{H^2}\dot\zeta\right)+\frac{\dot H}{H^2}\frac{\nabla_{G}^2\zeta}{a^2}=0\ .
\ee
Outside the sound horizon and assuming $c_s$ constant, we can simplify it to
\be
\dot\zeta=\frac{\dot H}{H^2}\frac{c_s^2 }{\left(\d_t\left(\frac{\dot H}{H^2}\right)+3\frac{\dot H}{H}\right)}\frac{\nabla_{G}^2\zeta}{a^2}\ .
\ee
Plugging back in (\ref{eq:constraint}), we can simplify the expression for $N$ and $N_i$ to be
\be
N\simeq 1\ ,\qquad  N_i=-\left(\frac{1}{H}+\frac{\dot H}{H^2}\frac{1}{3 H-\d_t\left(\dot H/H^2\right)}\right)\nabla_{G,i}\zeta\ .
\ee
At this point we proceed as in the main text. We Taylor expand $\zeta$ around the origin assuming spherical symmetry, we make an ansatz for the change of coordinates, and we impose the resulting metric to be in the Fermi form. After some straightforward algebra, we obtain for the change of coordinates
\bea\nonumber\label{eq:LocalGlobalMappingSimplzeta}
t_G&=& t_L-\frac{1}{2}\left[H(t_L)+\frac{\zeta_{,r_G r_G}}{a^2}\left(1-\frac{H \dot H}{-3H^4-2\dot H^2+H\ddot H}\right) \right]r_L^2 \ , \\ 
x_G^i &=&\frac{x_L^i}{a(t_L)}\left[1+\frac{H(t_L)^2}{4}
r_L^2\right]\left(1-\zeta(\vec 0)\right)\ ,
\eea
and for the metric
\bea\label{eq:metricPerturbedFermiSimplzeta}
&&ds^2=\\ \nonumber 
&&-\left\{1-\left[\dot H(t_L)+H(t_L)^2-\frac{\zeta_{,r_G r_G}}{a^2}\frac{1}{H^2\left(3H^4+2\dot H^2-H \ddot H\right)^2}\left(9 H^8 \dot H+9 H^6 \dot H^2+4\dot H^5-3 H^7 \ddot H\right.\right.\right.\\ \nonumber 
&&\left.\left.\left.-6 H^5\dot H\ddot H+2 H^3\dot H^2\ddot H-4 H\dot H^3\ddot
H+H^2\dot H\left(-2 \dot H^3+\ddot H^2\right)+H^4\left(12\dot H^3+\ddot
H^2-\dot H \dddot H\right)\right)\right]r_L^2\right\}dt_L^2\\ \nonumber
&&\qquad\ \  \,+\left\{1-\left(\frac{H(t_L)^2}{2}-\frac{\zeta_{,r_G
r_G}}{a(t_L)^2}\cdot\frac{H^2\dot H}{-3 H^4-2\dot H^2+H\ddot
H}\right)r_L^2\right\}d\vec x_L^2 \ . 
\eea
As expected, this metric has the same form as the Fermi patch of a closed FRW
Universe with
\bea\label{eq:localvariableszeta}
&&H_L(t_L)=H(t_L)+\frac{\zeta_{,r_G r_G}}{a(t_L)^2}\left(\frac{1}{H}+\frac{H \dot H}{3 H^4+2 \dot H^2- H\ddot H}\right)\ , \\ \nonumber
&&K_L=-\frac{2}{3}\nabla^2_G\zeta(\vec 0,t_G)\ ,
\eea
in agreement with what found in the Newtonian-gauge case.

\section{A Geometric Derivation of the Fermi Coordinates\label{app:geometrical}}
In this section we will describe how the Fermi coordinates can be constructed
from a geometric point of view.

 \begin{figure}[h!]
    \centering
        \includegraphics[width=.40\textwidth]{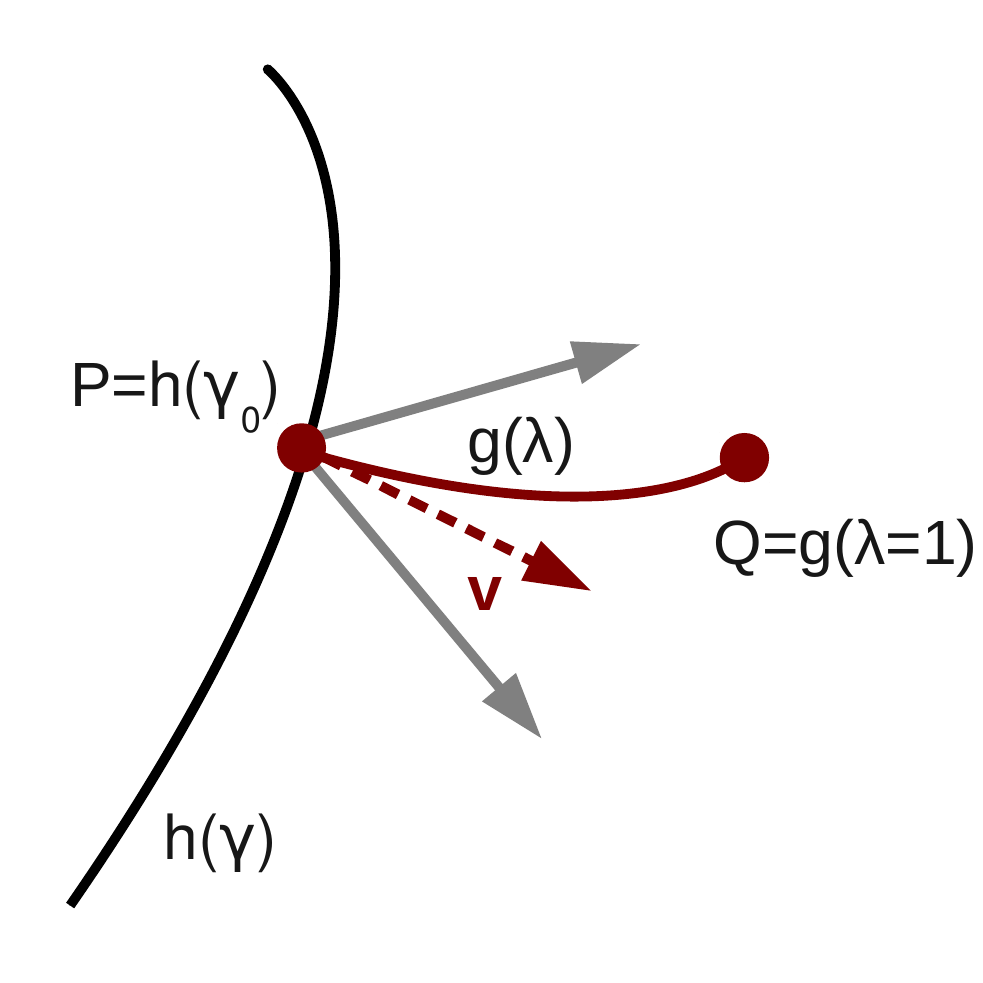}
    \caption{\sl Geometrical Construction of the Fermi Coordinates.}
    \label{fig:skizze}
\end{figure}

The starting point for the derivation will be a free falling observer moving
along a timelike geodesic $h(\gamma)$ in the background Universe
(fig.~\ref{fig:skizze}). His coordinate
axes
are described by an orthonormal set $(\vec{e}_0,\vec{e}_1,\vec{e}_2,\vec{e}_3)$
which is parallely transported along $h$. Thus if $\vec{e}_0$ is tangent to the
geodesic $h(\gamma)$ at its origin it will remain so for all values of the
affine parameter $\gamma$. Without loss of generality we can assume $\vec{e}_0$
to be
timelike and the $\vec{e}_i,\ i=1,2,3$ to be spacelike, and the geodesic to be the origin of the global coordinate frame
$x^i_G=0$.
\par
Now we consider a point $P=h(\gamma_0)$ on this geodesic.
Our goal is to describe the spacetime in a neighborhood $\mathcal{U}$ of
$P$ starting from the global metric at $P$. Any point $Q$ in the vicinity of $P$ can be
connected to $P$ with a geodesic $g(\lambda)$ that is perpendicular to the
tangent vector of $h$ at $P$, i.e., its tangent vector $\vec v$ at $P$ is a
linear
combination of the $\vec{e}_i$. The coefficients of this linear combination are
the Fermi coordinates and the time component of the Fermi coordinates is chosen
to be the proper time of the observer moving along $h$.
The point $Q$ can thus be fully described by the proper time $\tau$ of the
observer at $P$, the direction cosines $x^i$ and the length of the geodesic
$\lambda$ joining $Q$ with $P$. For simplicity we will normalize the direction
cosines such that the point $Q$ corresponds to $\lambda=1$. This
prescription is the natural extension of the flat space polar coordinates to
curved space. The observer points in a certain direction defined by
the direction cosines $x_L^i$ and then follows the geodesic defined
by the direction.
\par
The initial conditions for the geodesic connecting $P$ and $Q$ can thus
be summarized as
\begin{align}
x^i(\lambda=0)=0\ , && \tau(\lambda=0)=t_L\ , \\ \nonumber
\frac{dt_G}{d\lambda}\,\biggl.\biggr|_{\lambda=0}=0\ , &&
\frac{dx^i_G}{d\lambda}\, \biggl.\biggr|_{\lambda=0}=x^i_L
\vec{e}_i(t_0)\ .
\end{align}
The point $Q\in\mathcal{U}$ with Fermi coordinates $x_L^\mu$ is then
found by propagating along $g(\lambda)$ until~$\lambda=1$.
We now have to find the mapping between arbitrary coordinates $x^\mu$ and the
Fermi coordinates defining the geodesic $g(\lambda)$. This can be done by
solving the geodesic equation for~$g(\lambda)$
\be
\frac{d^2 x^\mu}{d\lambda^2}+\Gamma_{\alpha\beta}^\mu
\frac{dx^\alpha}{d\lambda}\frac{dx^\beta}{d\lambda}=0\ ,
\ee
perturbatively using the power law ansatz
\be
x^\mu(\lambda)=\alpha_0^\mu+\alpha_1^\mu\lambda+\alpha_2^\mu\lambda^2
+\alpha_3^\mu \lambda^3+\ldots.
\ee
The validity of this series is clearly limited as is the validity of the
Fermi coordinates themselves, which is obviously related to the curvature of the spacetime. 
The four vector formulation for the initial conditions stated above is
\begin{align}
\alpha_0^\mu=&(t_0,0,0,0),\\ \nonumber
\alpha_1^\mu=&\frac{\derd
x^\mu}{\derd \lambda}\bigg\vert_{\lambda=0}=x_L^i [\vec{e}_i]^\mu.
\end{align}
where $t_0$ is the coordinate time corresponding to $\gamma_0$.
The coefficients of the second and third order terms in the Taylor series follow
straightforwardly from the geodesic equation evaluated at $P$
\begin{align}
\alpha_2^\mu=&\frac{1}{2!}\frac{\derd^2
x^\mu}{\derd \lambda^2}\bigg\vert_{\lambda=0}=-\frac12\Gamma^\mu_{\gamma\nu}
\alpha_1^\gamma\alpha_1^\nu , \\ \nonumber
\alpha_3^\mu=&\frac{1}{6!}\frac{\derd^3
x^\mu}{\derd \lambda^3}\bigg\vert_{\lambda=0}=-\frac16\left(\frac{\partial
\Gamma^\mu_{\gamma\nu}}{\partial
x^\kappa} \alpha_1^\gamma\alpha_1^\nu\alpha_1^\kappa +4
\Gamma^\mu_{\gamma\nu}\alpha_1^\gamma\alpha_2^\nu\right),
\end{align}
where we already simplified using the initial conditions. In the following two
subsections we will describe the mapping for two specific cases: perturbed and
unperturbed FRW Universes.

\subsection{FRW}
We will now follow the above procedure for the homogeneous
Friedmann-Robertson-Walker metric
\be
\derd s^2=-\derd t_G^2+a(t_G)^2\frac{\derd
\vec{x}_G^2}{\left[1+\frac{1}{4}K\vec x_G^2\right]},
\ee
The vierbein associated to a comoving geodesic is
\begin{align}
[\vec{e}_0]^\mu=(1,0,0,0) \ ,&& [\vec{e}_1]^\mu=a^{-1}(0,1,0,0)\ ,\\ \nonumber
[\vec{e}_2]^\mu=a^{-1}(0,0,1,0)\ , && [\vec{e}_3]^\mu=a^{-1}(0,0,0,1)\ .
\end{align}
The linear coefficients in the geodesic expansion read
\be
\alpha_1^\mu=a(t_0)^{-1}(0,x_L,y_L,z_L)\ .
\ee
Hence, the first order spatial separation is $x^i_G\simeq x_L^i/a(t_{0})$ and
thus
$x_L^i$ is nothing but the physical separation of
$Q$ from $P$. Up to third order in the affine parameter we obtain
\begin{align}
t_G=&t_L-\frac{H \vec x_L^2}{2}\ , \\ \nonumber
x^i_G=&\frac{x^i_L}{a(t_L)}\left(1+\frac{H^2
\vec x_L^2}{3}\right)\ .
\end{align}
This leads to the following metric in Fermi Normal coordinates
\begin{align}
\derd
s^2=&-\Biggl[1-\left(\dot{H}(t_L)+H^2(t_L)\right)\vec{x}_L^2\Biggr]
dt_L^2\nonumber\\
&+\Biggl[
\delta_{ij}-\left(H^2(t_L)+\frac{K}{a^2}\right)\frac{\vec{x}_L^2
\delta_{ij}-x_L^i x_L^j}{3} \Biggr] \derd x_L^i \derd x_L^j
\end{align}
The above metric has non-zero off-diagonal contributions. The general
transformation to remove off diagonal terms can be derived considering the
metric in the old coordinates $\tilde{x}$
\be
\derd s^2=\tilde{A}\,\delta_{ij}\, \derd \tilde{x}^i\,
\derd \tilde{x}^j+\tilde{B}\,\tilde{x}_i
\tilde{x}_j\ \derd\tilde{x}^i\, \derd \tilde{x}^j\ ,
\ee
and new coordinates $x(\tilde{x})$
\be
\derd s^2=A\,\delta_{ij}\, \derd x^i\, \derd x^j\ .
\ee
Using the ansatz $\tilde{x}^i=x^i(1+\gamma x^2)$ we obtain the condition valid
at second order in $x$:
\bea
&&\gamma=-\frac{\tilde{B}}{4\tilde{A}}\\ \nonumber
&& A=\tilde{A}(1+2\gamma x^2)\ .
\eea
For the FRW case we have $\gamma=-H^2/12-K/(12a^2)$ and the time component up to
second
order is unaffected
\begin{align}
t_G=&t_L-\frac{H(t_L)}{2}\vec x_L^2\ ,\\ \nonumber
x^i_G=&\frac{x^i_L}{a(t_L)}\left(1+\frac{H(t_L)^2}{4}\vec
x_L^2\right)\ ,
\end{align}
finally leading to the following metric
\be
\derd
s^2=-\biggl[1-\left(\dot{H}(t_L)+H(t_L)^2\right)\vec{x}_L^2\biggr]
\derd t_L ^2+\left[1-\left(H(t_L)^2+\frac{K}{a(t_L)^2}\right)\frac{\vec
x_L^2}{2}\right] \derd \vec{x}_L^2\ ,
\ee
which has the desired form.

\subsection{Perturbed FRW}
Let us now consider a perturbed FRW Universe in Newtonian gauge
\be
\derd s^2=-\Bigl(1+2\Phi(t)\Bigr)\derd t^2+a^2(t)\Bigl(1-2\Psi(t)\Bigr)\derd
\vec x^2 \ .
\ee
We assume vanishing anisotropic stress leading to $\Phi=\Psi$.
The vierbein associated to the coordinate frame is
\begin{align}
[\vec{e}_0]^\mu=(1-\Phi,0,0,0)\ , && [\vec{e}_1]^\mu=a^{-1}(0,1+\Phi,0,0)\ , \\ \nonumber
[\vec{e}_2]^\mu=a^{-1}(0,0,1+\Phi,0)\ , && [\vec{e}_3]^\mu=a^{-1}(0,0,0,1+\Phi)\ .
\end{align}
We can now for simplicity expand the potentials around $P$
\be
\Phi(\vec x,t)
=\Phi(\vec 0, t)+\frac{1}{2}\frac{\partial^2 \Phi}{\partial r_G^2}\bigg\rvert_0
r_G^2
=\Phi(\vec 0,t)+\frac{1}{2}\Phi(\vec 0,t)_{,r_Gr_G} r_G^2\ ,
\ee
where we asumed spherical symmetry~\footnote{Note that
\[\nabla^2 \Phi=\frac{2}{r}\frac{\partial \Phi}{\partial r}+\frac{\partial^2
\Phi}{\partial r^2}=3\frac{\partial^2 \Phi}{\partial r^2} \]
where the last equality is true for a power law in $r$ assuming no linear
dependence in $r$.}
leading to
\be
\derd s^2=-\Bigl(1+2\Phi(\vec 0,t)+\Phi(\vec 0,t)_{,r_Gr_G}
r_G^2\Bigr)\derd t_G^2+a^2(t)\Bigl(1-2\Phi(\vec 0,t)-\Phi(\vec 0,t)_{,r_Gr_G}
r_G^2\Bigr)\derd\vec x^2_G,\label{eq:pertmet}\ .
\ee
There is no linear term in this expansion, because we require the potential to
be differentiable at $r=0$.
Let us proceed to find the Fermi coordinates. As noted above, the Fermi time is the proper time of the observer following the
central geodesic $h$
\be
t_L=\int_0^t \sqrt{-g_{00}}\, \derd t'=t+\int_0^t \Phi(\vec 0,t')\, \derd
t'\ .
\ee
The coordinate time at $P$ thus is
$t_0=t_L-\int_0^{t_L}\Phi(\vec 0,t')\
\derd t'$, where in the integral boundary $t_L=t_0$ at leading order in
$\Phi$.
This leads to the following expansion factors
\begin{align}
\alpha_0^\mu=&(t_0,0,0,0)\ , \\
\alpha_1^\mu=&\frac{1+\Phi(\vec
0,t_0)}{a(t_0)}\bigl(0,x_L,y_L,z_L\bigr)\ . \nonumber
\end{align}
At this point, simple algebra as shown in the former section leads to the same relationship among the coordinates as in~\eqref{eq:LocalGlobalMappingSimpl} and to the same Fermi metric as in~\eqref{eq:metricPerturbedFermiSimpl}.
\subsection{Local Expansion Factor\label{app:local_expansion}}

With the aim of giving very specific recipe for running simulations given a
certain long wavelength fluctuation, we provide some more specific relations.
Some expressions can be simplified by noticing that the potential in Newtonian
gauge and the density perturbation in the comoving gauge are
related by (see Appendix \ref{app:symbols})
\be
\nabla^2 \Phi(\vec x,t)=4\pi G a^2 \bar \rho\, \delta^{(com)}_l(\vec x,t)\ .
\label{eq:poisssc}
\ee
Note that this equation is exact, even on horizon scales. We define the growth
factor in comoving synchronous gauge as
$\delta_l^{(com)}(t)=D(t)\delta_{l,0}^{(com)}$, which is normalised to unity at
present time. We also define the logarithmic
growth factor $f(a)=\derd \ln{D}/\derd \ln{a}$. From
eq.~\eqref{eq:poisssc}, we define the growth factor of the Newtonian potential $D$ as follows
$\Phi(t)=D(t) \Phi_0/a(t)$, where $\Phi_0$ is the present day value.
From the linear growth and eq.~\eqref{eq:poisssc} it follows
\be
\Phi(\vec 0,t)_{,t\, r_Gr_G}=H\Phi(\vec 0,t)_{,r_Gr_G}(f-1)\ . \label{eq:phidot}
\ee
Using the constancy of $\zeta$ we have shown that the value of the metric
perturbation at the origin is irrelevant for the local expansion. Thus it only remains to derive the
rescaling of the expansion factor corresponding to the effective local Hubble
rate. Starting from (\ref{eq:localvariables}), which, by defining $H_L(t)=H_G(t)+\delta H(t)$, gives
\be
\delta H=\frac{1}{a_G^2(t)}\frac{H_G(t)}{\dot{H}_G
(t)} \left(\Phi(\vec 0,t)+\frac{\Phi(\vec
0,t)_{,t}}{H}\right)_{,r_Gr_G}\ ,
\ee
we can find the corresponding rescaling for the expansion
factor using the ansatz
\be\label{eq:a_rescaling}
a_L(t)=a_G(t)\left(1+\delta a(t)_{rel}\right)\ ,
\ee
where $\delta a_{rel}$ has to satisfy the following:
\bea
&&\dot{\delta a}_{rel}(t)=\frac{1}{3a_G^2(t)H_G(t)}\nabla_G^2\left(\Phi(\vec
0,t)+\zeta(\vec
0,t)\right) \ ,\qquad\\ \nonumber
&& \Rightarrow\qquad \delta a_{rel}(t)=\int^t_0 dt'
\frac{1}{3a_G^2(t')H_G(t')}\nabla_G^2\left(\Phi(\vec 0,t')+\zeta(\vec
0,t')\right)\ ,
\eea
where we have chosen the constant so that the two scale factors agree at early times. $\delta a(t)_{rel}$~can be numerically integrated from the transfer functions for any given cosmology.

\par
Finally, the Friedmann equations in the Fermi frame read as
\be
H^2_L=\frac{8\pi
G}{3}\bar{\rho}_L+\frac{8\pi G}{3}\bar\rho_{DE}-\frac{K}{a^2_L}\ ,
\label{eq:locfried1}
\ee
and
\be
\frac{\ddot{a}_L}{a_L}=-\frac{4\pi
G}{3}\bar{\rho}_L+\frac{8\pi G}{3}\bar\rho_{DE}\ ,
\label{eq:locfried2}
\ee
where $\bar{\rho}_L$ is the local mean matter density.
From the rescaling between the global and local Hubble rate we can derive the
rescaling of the local mean density
\be
\frac{H_L(t)^2+\frac{K}{a(t)^2}-\frac{8\pi
G}{3}\bar\rho_{DE}(t)}{H_G(t)^2-\frac{8\pi
G}{3}\bar\rho_{DE}(t) } =\frac{\bar{\rho}_L(t)}{\bar {\rho}_G(t) }\ ,
\ee
leading to
\be\label{eq:rholocal}
\bar{\rho}_L(t)=\bar{\rho}_G(t)+\frac{3\Phi(\vec 0,t)_{,r_Gr_G}}{4\pi G
a^2(t)}=\bar{\rho}_G(t)\bigl(1+\delta_l^{(com)}(t)\bigr)\ .
\ee
This relation can be intuitively understood in the Newtonian context: the long
wavelength density just rescales the local mean density. This relationship gets
upgraded to the relativistic setup by using the comoving gauge
overdensity.
\par
For definiteness we give also the closed form expressions for 
the expansion and Hubble rate in a $\Lambda$CDM background
\begin{align}
H_L(t)=&H_G(t)\left(1-\frac{f(t)\Phi(\vec 0,t)_{,r_Gr_G}}{4\pi G a^2(t)
\bar\rho_G(t)} \right)=H_G(t)\left(1-\frac{1}{3} f(t)\delta_l^{(com)}(t)\right)\
, \\ \nonumber
a_L(t)=&a_G(t)\left(1-\frac{\Phi(\vec 0,t)_{,r_Gr_G}}{4\pi G a^2(t)
\bar\rho_G(t)} \right)=a_G(t)\left(1-\frac{1}{3}\deltal^{(com)}(t)\right).
\end{align}
\subsection{Local Density Parameters\label{app:local_parameters}}
The time evolution of the local patch is determined by the local
Friedmann eqns.~\eqref{eq:locfried1} and~\eqref{eq:locfried2}, which are
parametrized by the effective local density parameters.
We will now provide the explicit mapping from the global to the local
cosmological parameters that are needed for simulations. We specialize to $\Lambda$CDM for simplicity, though, as we stressed, our approach applies also to clustering dark energy.
From the relationship between $H_G$ and $H_L$ and from the definition of of $K$
given in (\ref{eq:localvariables}), we have
\bea
&&\Omega_{K,L}(t_L)=-\frac{K}{a_L(t_L)^2 H_L(t_L)^2}=\frac{2}{3}\frac{1}{
a_L(t_L)^2 H_L(t_L)^2}\nabla^2_G\zeta(\vec x_G)\ , \\ \nonumber
&&\Omega_{\Lambda,L}(t_L)=\frac{\Lambda}{3H_{L}^2}\ , \qquad\qquad \Omega_{m,L}(t_L)=1-\Omega_K(t_L)-\Omega_\Lambda(t_L)\ .
\eea
It is convenient to normalize the cosmological parameters at $a=1$
(for us $a_L=1$), which leads to
\bea
&&\Omega_{K,L,0}=-\frac{K}{H_{L,0}^2}=\frac{2}{3}\frac{1}{
H_{L,0}^2}\nabla^2_G\zeta(\vec x_G)\ , \\ \nonumber
&&\Omega_{\Lambda,L,0}=\frac{\Lambda}{3H_{L,0}^2}\ , \qquad \qquad\Omega_{m,L,0}=1-\Omega_{K,0}-\Omega_{\Lambda,0}\ ,
\eea
where the subscript $_0$ stays for evaluating the quantity when $a_L=1$. In
order to be able to use the above formulas, we simply need to find the time
$t_L$ at which $a_L=1$. This can be found by solving eq.~(\ref{eq:a_rescaling})
with $a_L=1$ to identify $t_{L,0}$. From there, by plugging into
(\ref{eq:localvariables}) we get $H_{L,0}$. The former expressions can be further simplified to give
\begin{align}
\Omega_{m,L,0}=&\frac{8\pi G
\bar{\rho}_L}{3H_{L,0}^2}=\omnot\left[1
-\frac{2}{3}\frac{ \nabla_G^2 \zeta(\vec 0)}{H_{G,0}^2}\right]=\omnot\left[1
+\left(\omnot+\frac{2}{3}f_0\right)\delta_{l,0}^{(com)}\right]
\\ \nonumber
\Omega_{K,L,0}=&-\frac{K}{H_{L,0}^2}=\frac{2}{3}\frac{ \nabla_G^2
\zeta(\vec 0)}{H_{G,0}^2}=-\left(\omnot+\frac{2}{3}
f_0\right)\delta_{l,0}^{(com)} \\ \nonumber
\Omega_{\Lambda,L,0}=&\frac{\Lambda}{3 H_{L,0}^2}
=\left(1-\omnot\right)\left [
1-\frac{2}{3}\frac{ \nabla^2_G \zeta(\vec
0)}{H_{G,0}^2}\right]=\left(1-\omnot\right)\left [
1+\left(\omnot+\frac{2}{3}f_0\right)\delta_{l,0}^{(com)}\right]\ .
\end{align}
The local Hubble rate at $a_L=1$ is given by
\be
H_{L,0}=H_{G,0}\left(1+\frac{1}{3}\frac{ \nabla_G^2
\zeta(\vec
0)}{H_{G,0}^2}\right)=H_{G,0}\left[1-\frac{1}{2}\left(\Omega_{m,0}+\frac{2}{3}
f_0
\right)\delta_{l,0}^{(com)}\right]\ .
\ee
As an example we consider the WMAP5 Flat $\Lambda$CDM cosmology
with matter density parameter $\omnot=0.28$ and Hubble constant $H_{G,0}=70\,
\text{km}\, \text{s}^{-1}\, \text{Mpc}^{-1}$. For a long wavelength amplitude of
$\delta_{l,0}^{(com)}=0.1$ corresponding to $\nabla_G^2
\zeta/H_{G,0}^2=0.89$ we obtain
\begin{align}
 \Omega_{m,L,0}=0.30\ , && \Omega_{K,L,0}=-0.06\ , &&
\Omega_{\Lambda,L,0}=0.76 && h_{L,0}=0.68\ ,
\end{align}
where we wrote the Hubble constant in terms of $h_{L}$ as $H_{L,0}=100h_L\,
\text{km}\, \text{s}^{-1}\, \text{Mpc}^{-1}$.\\
In fig.~\ref{fig:localexpansion} we show the time dependence of the effective
local expansion history. At early times the curvature is negligible and the
effective local Universe approaches the flat background Universe. At late
times, the cosmological constant dominates and thus the contribution of matter
and curvature to the energy budget becomes irrelevant. 
\begin{figure}[t]
\includegraphics[width=0.5\textwidth]{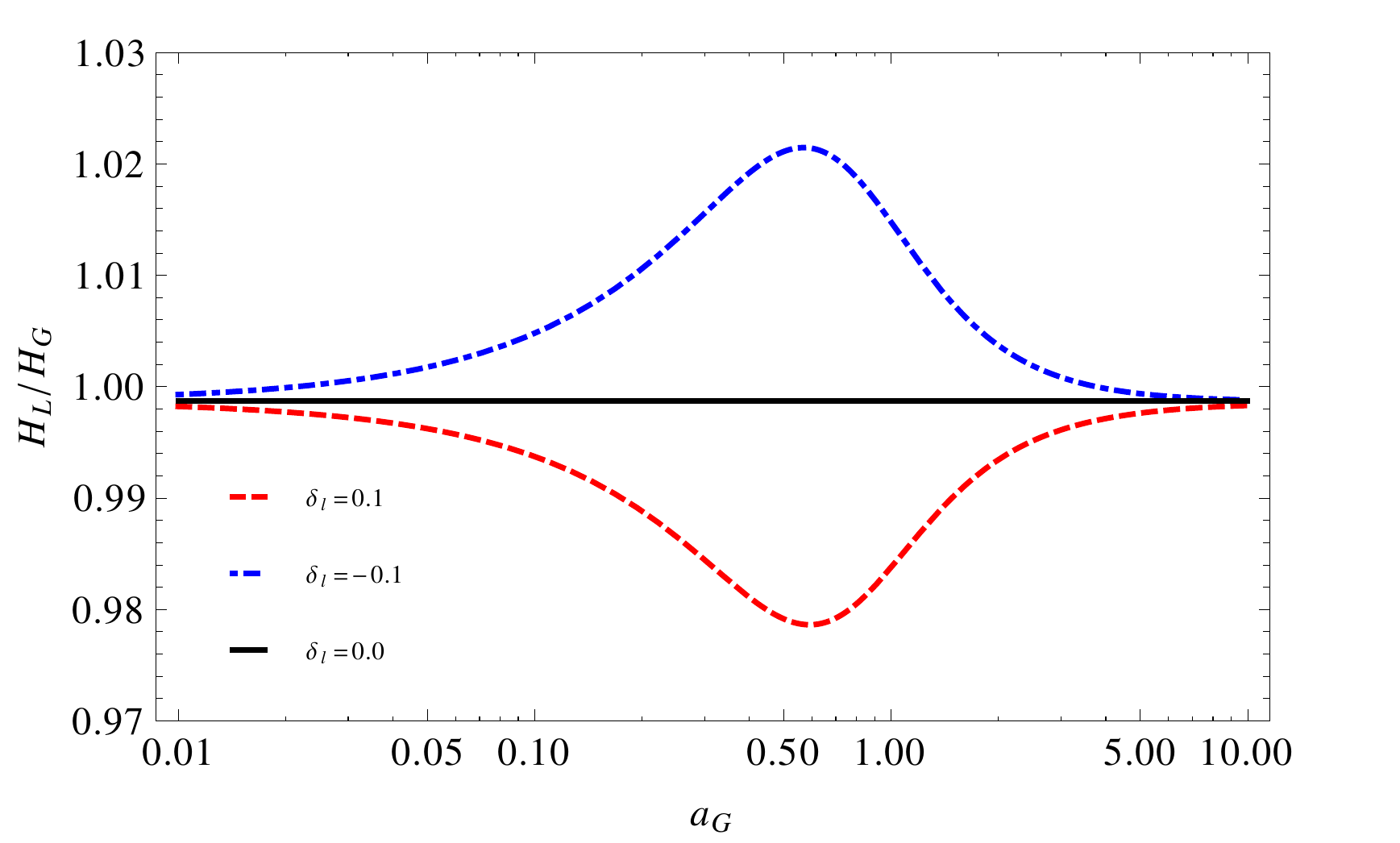}
\includegraphics[width=0.5\textwidth]{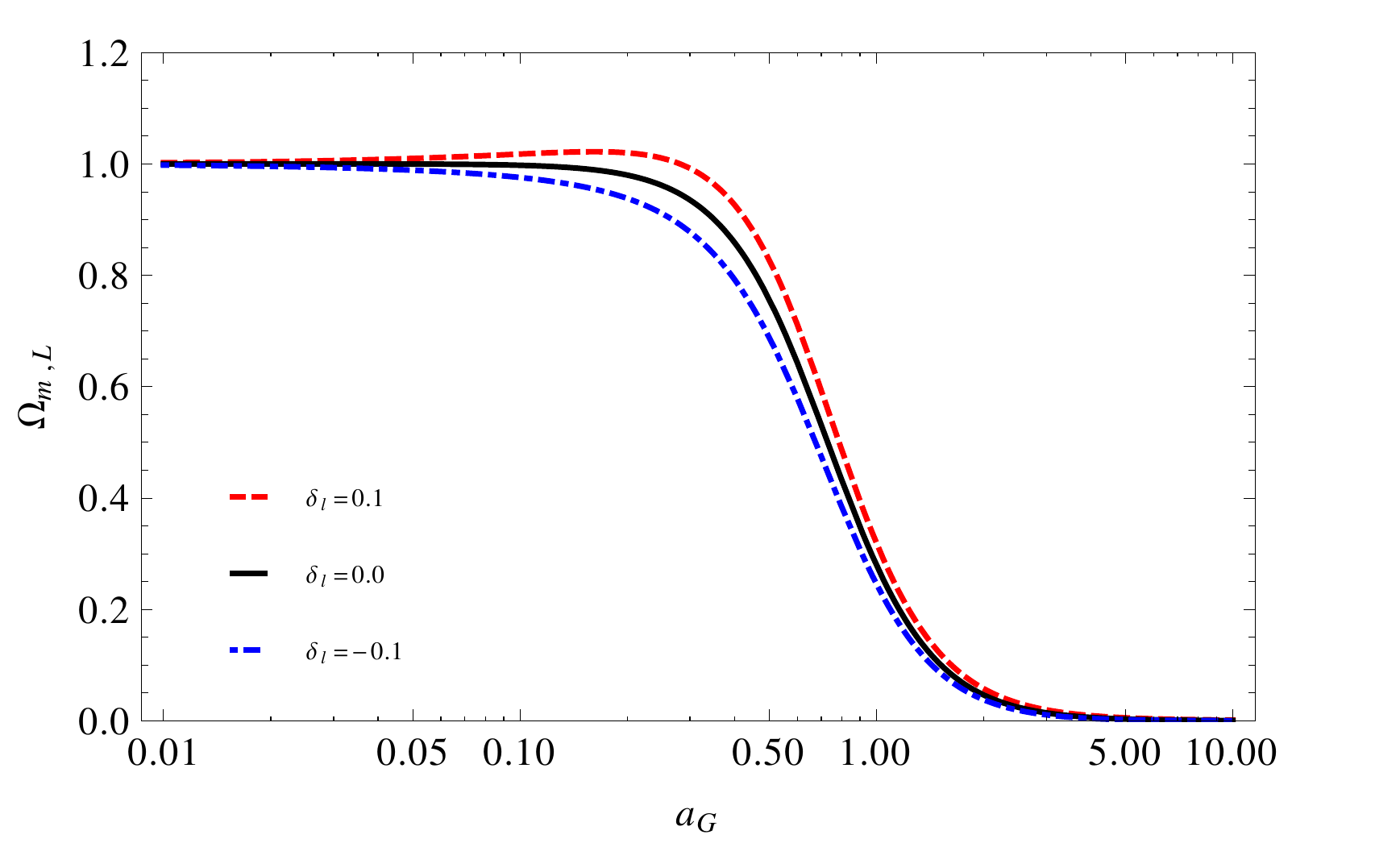}\\
\includegraphics[width=0.5\textwidth]{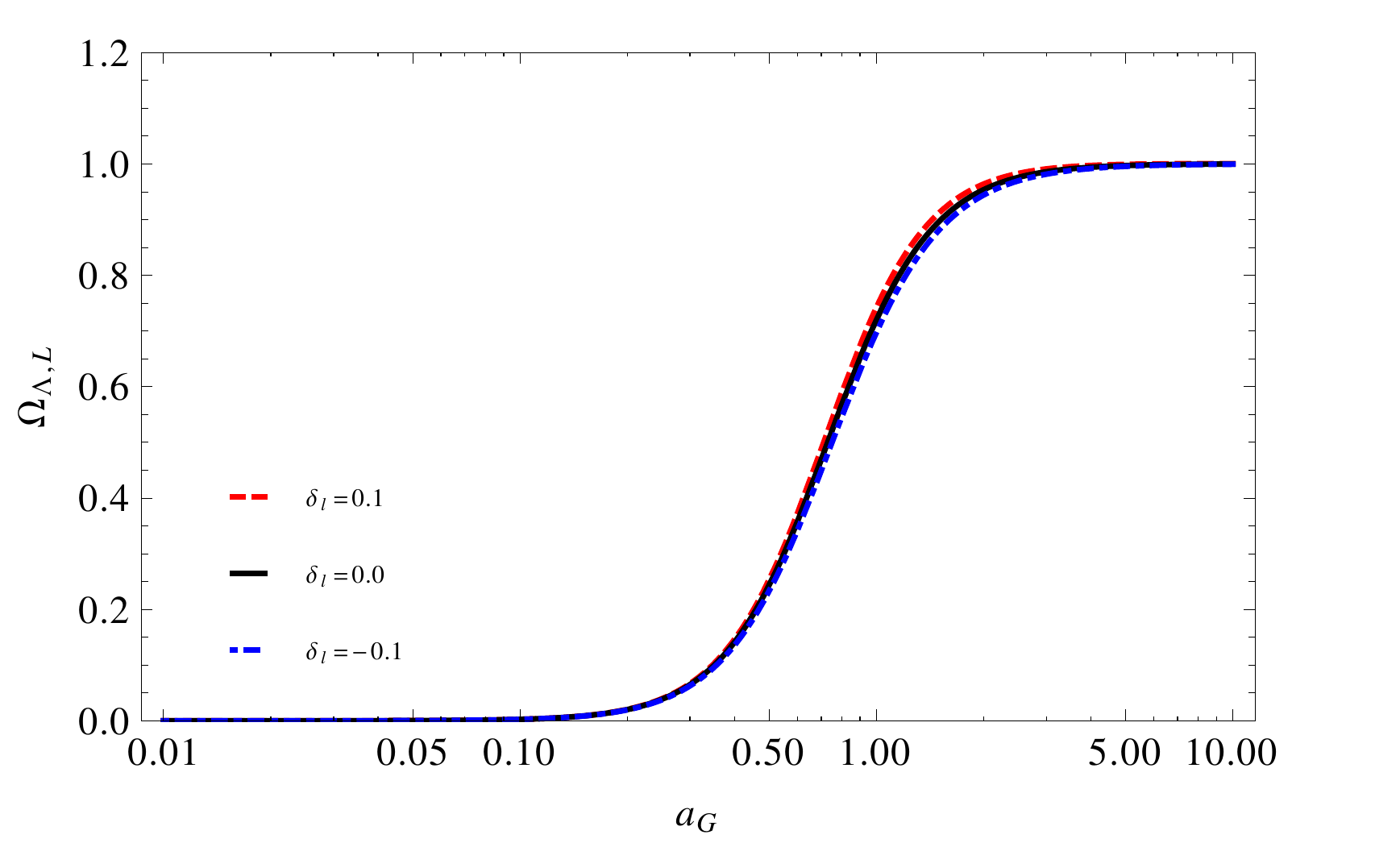}
\includegraphics[width=0.5\textwidth]{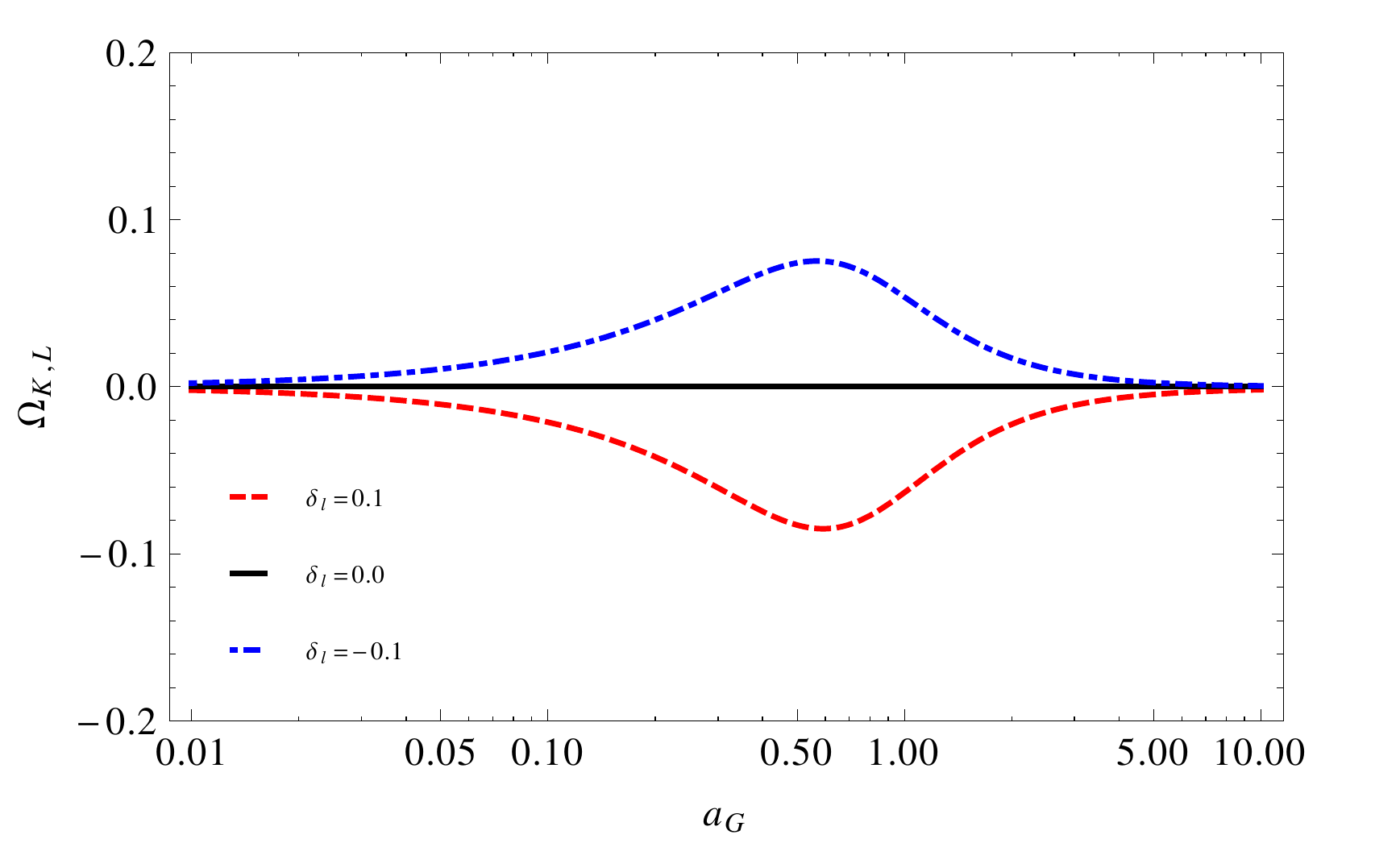}
\caption{\sl Time dependence of the local expansion history as a function of the
global expansion factor. In all panels the
solid black line represents the flat background model, whereas the red dashed
and blue dash-dotted lines represent an over- or underdense region.
\emph{Top left: } Ratio of the local and global Hubble rate.
\emph{Top right: } Local matter density parameter.
\emph{Bottom left: } Local cosmological constant density parameter.
\emph{Bottom right: } Local curvature density parameter.
}
\label{fig:localexpansion}
\end{figure}
\section{Fermi Coordinates for Plane-Wave Perturbed FRW
Universe}\label{app:Fermi-wave}

Let us assume we start in Newtonian gauge with $\Phi(\vec x_G,t_g)$ being a plane wave with wave-number that we can take without loss of generality in the $x$-direction
\be
\Phi(\vec x_G,t_G)=\Phi_0\, e^{i k x_G}\ .
\ee
We can find the Fermi coordinates around the origin by working as in the main text and assume a change of coordinates valid at cubic order in the spatial distance of the form:
\bea
&&t_G= t_L-\frac{1}{2}H(t_L)r_L^2-\int_0^{t_L} \Phi(\vec 0,t')dt'+g_1(t_L) r_L^2+\\ \nonumber
&&\qquad+g_{a,1;}{}_j(t_L) x_L^j+g_{a,2;}{}_{j,k}(t_L) x_L^jx_L^k \ , \\
\nonumber
&& x_G^i =\frac{x_L^i}{a(t_L)}\left(1+\frac{1}{4}H(t_L)^2r_L^2+f_1(t_L)+f_2(t_L)r_L^2\right)\\ 
\nonumber&&\qquad +f_{a,tr}^i(t_L)+f_{a,0;}^i{}_j(t_L)
x_L^j+f_{a,1;}^i{}_{j,k}(t_L) x_L^jx_L^k+f_{a,2;}^i{}_{j,k,l}(t_L)
x_L^jx_L^kx_L^l\ .
\eea
This represents the most general change of coordinates around $\vec x_G=\vec 0$
at cubic order in the distance from the origin, and it is a straightforward
generalization of (\ref{eq:change}). The subscript $a$ represents the fact that
those functions are zero in the limit of isotropic perturbations. By imposing
the metric in the new local coordinates to be of the Fermi form, we can proceed
and identify the unknown functions. We skip the details associated to
straightforward algebra, and just quote the final result. In order to limit the
size of the expressions, we simply quote the simple expressions that are
obtained after we restrict to the case of constant $\zeta$. 

Under the following change of coordinates
\bea\nonumber\label{eq:LocalGlobalMappingSimplwave}
&&t_G= t_L-\int_0^{t_L} \Phi(\vec 0,t_L)dt'-\frac{1}{2}H(t_L)\left(1-\Phi(\vec 0,t_L)\right) r_L^2\\ \nonumber
&& \quad\qquad+ \frac{x_L^1}{a(t_L) H(t_L)} \left. \nabla_{G,1}\left[\Phi(\vec
x_G,t_L)+\zeta(\vec x_G,t_L)\right]\right|_{\vec x_G=0} \ , \\ 
&&x_G^i=\frac{x_L^i}{a(t_L)}\left\{\left[1+\frac{H(t_L)^2}{4} r_L^2\right]\left(1-\zeta(\vec 0,t_L)\right)-\frac{x_L^1}{a(t_L)}\nabla_{G,1}\zeta(\vec x_G,t_L)\right\}\\ \nonumber
&&\quad\qquad+\delta_{i,1}\left[\left.\nabla_{G,1}\int_0^{t_L}dt'
\frac{\Phi(\vec x_G,t_L')+\zeta(\vec x_G,t_L')}{H(t_L')a(t_L')^2}\right|_{\vec
x_G=0}-\frac{1}{2}\frac{r_L^2}{a(t_L)^2}\nabla_{G,1}\Phi(\vec x_G,t_L)\right. \\
\nonumber
&&\left.\quad\qquad+\frac{(x_L^1){}^3}{6 a(t_L)^3}\left.\nabla_{G,11}^2\Phi(\vec x_G,t_L)\right|_{\vec x_G=0}\right]\ ,
\eea
we obtain the following metric components:
\bea
&& g_{00}=-1+\left(H(t_L)^2+\dot H(t_L)\right)r_L^2-\frac{(x_L^1)^2}{a(t_L)^2}\left.\nabla^2_{G,11}\Phi(\vec x_G,t_L)\right|_{\vec x_G=0}\ , \\ \nonumber
&& g_{0i}=\frac{H(t_L)}{4a(t_L)}\left(2 x_L^i x_L^1+\delta_{i1}r_L^2\right) \left.\nabla_{G,1}\left[\zeta(\vec x_G,t_L)+3\Phi(\vec x_G,t_L)\right]\right|_{\vec x_G}\ ,\\ 
\nonumber
&& g_{ij}=\delta_{ij}\left(1-\frac{r_L^2}{2}H(t_L)^2-\frac{(x^1_L)^2}{a(t_L)^2}\left.\nabla_{G,11}\Phi(\vec x_G,t_L)\right|_{\vec x_G}\right)+\delta_{i1}\delta_{j1}\frac{(x^1_L)^2}{a(t_L)^2}\left.\nabla_{G,11}^2\Phi(\vec x_G,t_L)\right|_{\vec x_G}\ ,
\eea
We see that in the anisotropic case, the metric has non-vanishing $0i$
components at order $r_L^2$. This form of the metric is important if we are
interested in evaluating the bias for a non-scalar quantity, for which case the
problem can not be reduced to the spherically symmetric case.

\section{Growth in Presence of a Long Mode}\label{app:collobjects}


In the main text, we derived a change of coordinates that is valid in a small
region around a given time-like geodesic and that allowed us to describe the
effect of a long wavelength fluctuation effectively as a local closed FRW
Universe. This change of coordinates is valid at linear order in the long mode
and at any
order in the short wavelength perturbations. In the main text we focus on
collapsed objects, as our main interest is extracting information about
halo bias. Therefore we follow the short scale power well into the
non-linear regime. On the other hand, the mapping can also be used to
analytically examine the coupling of linear short wavelength modes to long
wavelength modes while the short modes are still in the quasi-linear regime. In this regime, we are now going to explicitly compare results
derived in our formalism to the ones obtained in standard perturbation theory. Since standard
perturbation theory is performed in the Newtonian limit ($k/aH\gg 1$), we will
adopt the simplifying assumption that the long mode is sufficiently far inside
the horizon that the Newtonian approximation holds also for the long mode. In this limit we can for
example neglect $\Phi \ll \nabla^2\Phi/(a^2 H^2)$~\footnote{We stress that we
perform this approximation just in this appendix to make contact with former
literature, but we do not do this same approximation in the main text, where the
derivation is performed in full GR.}. We will also restrict ourselves to
the Einstein de-Sitter (EDS) Universe for simplicity.

As shown in the main text, the effect of a long wavelength mode on the
local dynamics can be ascribed to a non-vanishing spatial curvature $K$ in a
fictious closed global FRW Universe. The curvature parameter $\Omega_K$ and
the curvature $K$ are
related by
\be
\Omega_K=-\frac{K}{a^2 H^2}.
\ee
Let us begin to investigate the growth of short scale fluctuations in a closed
FRW Universe. The linear growth equation for the short wavelength matter density perturbations
reads as
\be
\ddot{\delta}_s+2H\dot{\delta}_s-4\pi G
\bar{\rho}\delta_s=0\ .
\ee
This equation is solved by the linearly growing modes
$\delta_s(t)=D(t)\delta_\text{s,0}$:
\be
D(t)=\frac{5}{2}\Omega_{m}H_0^2
H(t)\int_0^{a(t)}\frac{d\tilde{a}}{\left[\tilde{a }
H(\tilde{a})\right]^3 }\ ,
\ee
which in EDS simplifies to $D(t)=a(t)$, where we use the subscript $_0$ to indicate present time and where we have normalized $a_0=1$. We will now look at the relation
between the linear growth in a globally flat Universe and the effective local curved
Universe.
Using $\dot{H}(t)=-3H^2(t)/2$ in EDS, we obtain for the local
effective curvature in terms of the long wavelength density perturbation
\be
K=2\left[\Phi(\vec
0,t_L)-\frac{H_G^2(t_L)}{\dot{H}_G(t_L)}\left(\Phi(\vec
0,t_L)+\frac{\Phi(\vec
0,t_L)_{,t_L}}{H_G(t_L)}
\right)\right ]_{,r_Gr_G}=\frac{5}{3}H_G^2a_G^3\delta_{l,0}\ .
\ee
Here we have inserted a subscript $_l$ to $\delta$ to make it more explicit that
it represents a long wavelength fluctuation.
The growth now depends on the effective curvature in two ways. First, the growth
in overdense regions is enhanced by a factor of $20 \delta_l/21$.
Furthermore, from~(\ref{eq:a_rescaling}) we obtain for the relation between local
and global expansion $a_L(t_L)=(1-\delta_l/3)a_G(t_L)$, and, at a given fixed proper time, thus we have to
evaluate the local growth at an earlier (later) scale-factor for
overdense (underdense)
regions. This partially cancels the first dependence. Adding both
contributions, the
derivative of the growth rate with respect to the long wavelength density reads
as
\be
\left.\frac{\d D}{\d \delta_{l,0}}\right|_{a_G}=
\frac{\partial D}{\partial \Omega_{K}}\bigg\lvert_{a_L}
\left.\frac{\d \Omega_K}{\d \delta_{l,0}}\right|_{a_G}
+\frac{\partial D}{\partial a_L}\bigg\lvert_{\Omega_K}
\left.\frac{\d a_L}{\d\delta_{l,0}}\right|_{a_G}=\frac{20}{21}a_G^2-\frac{1}{3}a_G^2=\frac{13}{
21}a_G^2\ ,
\ee
Thus, we finally have with $\delta_l(t)=a_G\delta_{l,0}$:
\be
D(\delta_l\neq0)=D_0+\frac{\d D}{\d
\delta_{l,0}}\bigg\rvert_{0}\delta_{l,0}
=a_G\left(1+\frac{13}{21}\delta_l\right)
\ee
It turns out that the coupling strength of $13/21$ is a particlular property of
the Einstein-de-Sitter Universe. In a more general $\Lambda$CDM Universe the
coupling is less strong. Thus we will write the enhanced growth generally as
$D=D_0(1+\beta\delta_l)$ in the following.
From the rescaling between the global and local Hubble rate we can derive the
rescaling of the local mean density
\be
\frac{H_L^2+K/a^2}{H_G^2}=\frac{\bar{\rho}_L}{\bar{\rho}
_G}
\ee
leading to
\be
\bar{\rho}_L=\bar{\rho}_G(1+\delta_l)
\ee
Since $\rho$ is a scalar, local and global density agree
${\rho}_G(x)=\rho_L(x)$ when evaluated at the same physical point.
Given that in this approximation we are neglecting the difference between $t_L$ and $t_G$,we have
\be
 \delta_L(\vec x)=\frac{\rho(\vec x)}{\bar{\rho}_L}-1\ , \quad
\delta_G(\vec x)=\frac{\rho(\vec x)}{\bar{\rho}_G}-1 \quad\Rightarrow\quad
\delta_G=(1-\delta_l)(1+\delta_L)-1\ .
\ee
Manipulating the last expression, we obtain
\be
\delta_G(x)=\delta_{L,0}(1+\beta \delta_l)(1+\delta_l)+\delta_l=\delta_{L,0}[1+(1+\beta) \delta_l]+\delta_l=\delta_{L,0}\left(1+\frac{34}{21} \delta_l\right)+\delta_l\ ,
\ee
where in the last step we have assumed EDS Universe.
Here, we accounted both for the excess growth in the local frame and for the
rescaling of the local mean density, with respect to which the local
overdensity is defined.
The three point function between long and short modes thus reads
\be
\la\delta_\text{G,s}(\vec x) \delta_\text{G,s}(\vec x)
\delta_l(\vec x)\ra=2\times\frac{34}{21}\sigma_s^2 P_l(k)\ .
\label{eq:ourcoupling}
\ee
\subsection{Correlators between Long and Short Modes}
The coupling between long and short modes in the Newtonian regime can also be
examined using perturbation theory (for a review see
\cite{Bernardeau:2001qr}).
Standard perturbation theory solves the Newtonian fluid equations using a
perturbative expansion in matter density and velocity divergence. In an
Einstein de Sitter Universe, the second order contribution to the matter density
field can be calculated as
\be
\delta^{(2)}(\vec k)=\int \frac{d^3q}{(2\pi)^3}F_2(\vec q,\vec k-\vec
q)\;\delta^{(1)}(\vec q)\;\delta^{(1)}(\vec k -\vec q)
\label{eq:sptmodecoupling}
\ee
where $\delta^{(1)}$ is the linearly evolved primordial density field and the
second
order mode coupling kernel is defined as
\be
F_2(\vec k_1,\vec k_2)=\frac{5}{7}+\frac{1}{2}\frac{\vec k_1\cdot \vec
k_2 }{k_1
k_2}\left(\frac{k_2}{k_1}+\frac{k_1}{k_2}\right)+\frac{2}{7}\frac{\left(\vec k_1
\cdot \vec k_2\right)^2}{k_1^2 k_2^2}\ .
\ee
We can now apply \eqref{eq:sptmodecoupling} to the case where we have a
Universe with short modes $\delta_s(\vec k_s)$ and a spherical
symmetric monochromatic long mode $\delta_l(
k_l)$. In this case we get for the matter field up to second order
\begin{align}
\delta^{(2)}_s(\vec k_s)=&\int\frac{\derd
\Omega_l}{4\pi} F_2(\vec k_s,\vec
k_l)\delta^{(1)}_s(\vec k_s)\delta^{(1)}_l(
k_l)=\delta^{(1)}_s(\vec k_s)\left(1+\frac{34}{21}\delta^{(1)}_l(
k_l)\right)\ ,
\end{align}
where we neglected the coupling of the short and long modes with themselves.
\par
The skewness of the density field at second order is
\be
\la\delta(\vec x)^3\ra=6\int \dqc \int \dqcp P(q)P(q')F_2(\vec q, \vec
q')=3 \times \frac{34}{21}\sigma^4\ .
\ee
The prefactor $3$ arises from the fact that all three density fields in
$\la\delta^3\ra$ can be expanded to second order.
For the correlator between short and long modes we obtain
\be
\la \delta_s(\vec x)\delta_s(\vec x)\delta_l(\vec
x)\ra=4\int \dqc \int \dqcp P_s(q)P_l(q')F_2(\vec q, \vec
q')=2\times \frac{34}{21}\sigma_s^2P(k_l)\ ,
\ee
where we assumed $P_l(\vec q)=(2\pi)^3 \delta^\text{(D)}(\vec q-\vec
k)P(\vec k)$, $k_l$ is the long wavelength and the prefactor 2 arises from the
fact that now only two of the three fields can be expanded to second order.
This is in perfect agreement with our result in~\eqref{eq:ourcoupling}.

\section{Spherical Collapse Dynamics\label{app:curved_spherical_collapse}}
The collapse of a dark matter halo can be calculated considering a spherical
overdensity within an otherwise homogeneous background Universe. In the standard
calculation the background is assumed to be a flat matter-only Universe (aka
Einstein-de-Sitter Universe). After reviewing the standard spherical collapse
dynamics we extend the calculation to the case where the background Universe is
curved. As we argued in the main text, this corresponds to the collapse in
the presence of a long-wavelength mode. This procedure will offer us a way to match
our General Relativistic definition of the bias with the standard Newtonian
definition.

\subsection{Collapse in Flat FRW}
According to Birkhoff's theorem, a spherically symmetric overdense region
evolves as a closed FRW Universe, whose Friedmann equation reads as
\be
H_C^2=\left(\frac{\dot{a}_C}{a_C}\right)^2=\frac{8\pi G \rho_C}{3
a_C^3}-\frac{K_C}{a_C^2}\ ,
\ee
where the subscript $C$ is used to refer to the collapsing region. This
collapsing region typically has the size of a dark matter halo and should not be
mistaken for the local patch described in the main part of this paper, which can
contain many of these collapsing regions.
The time evolution of the scale factor of the closed patch can be parametrized
by the cycloid solution
\be
a_C=A_C \left(1-\cos{\theta}\right)\ ,\qquad
t=B_C\left(\theta-\sin{\theta}\right)\ , \quad {\rm with}\quad
\theta\in[0,2\pi]\ ,\label{eq:collpara}
\ee
where we defined 
\be
A_C=\frac{4\pi G \rho_C}{3 K_C} \ , \quad{\rm and}\quad
B_C=\frac{4\pi G \rho_C}{3 K_C^{3/2}}\ .
\ee
The spherical overdense region described by this parametrization expands until
$\theta=\pi$, then it turns around to collapse at $\theta=2\pi$, corresponding
to the collapse time $t_{coll}=2\pi B_C$. Formally the expansion at the collapse
time is zero, but physically one expects the region to form a virialized object
at some time between turnaround and collapse.
At early times $\theta \ll 1$ the parametric solution can be expanded as
\begin{align}
a_C=&A_C \frac{\theta^2}{2}\left(1-\frac{\theta ^2}{12}+\frac{\theta
^4}{360}-\frac{\theta
^6}{20160}+\ldots\right)\ ,\\
t=&B_C\frac{\theta^3}{6}\left(1-\frac{\theta ^2}{20}+\frac{\theta
^4}{840}-\frac{\theta
^6}{60480}+\ldots\right)\ .
\end{align}
Solving the above equations consistently up to order $\mathcal{O}(\theta^4)$
one obtains
\be
a_C=A_C\frac{6^{2/3}}{2}\left(\frac{t}{B_C}\right)^{2/3}
\left[1-\frac{6^{2/3}}{20}\left(\frac{t}{B_C}\right)^{2/3}
\right]\ .
\ee
The linear overdensity of the closed Universe collapsing at $t_{coll}$ is then
given by the fractional deviation between the local and the background volume
(described here by the respective expansion factors)
\be
\delta(t;t_{coll})=\frac{a_B^3}{a_C^3}-1=\frac{3}{5}\left(\frac{3\pi}{2}
\right)^ { 2/3 } \left(\frac{t}{t_{coll}}\right)^{ 2/3
}=\frac{3}{5}\left(\frac{3\pi}{2}\right)^{2/3}
\frac{1+z_{coll}}{1+z(t)}\ ,
\ee
where we have used that the matter-only background Universe evolves according to
$a_B\propto t^{2/3}$.
Finally, one obtains the critical density for collapse at $z_{coll}$, linearly
extrapolated to the present time $z(t_0)=0$
\be
\delta_c(z_{coll})=\frac{3}{5}\left(\frac{3\pi}{2}\right)^{
2/3}\left(1+z_{coll}\right)\approx 1.686 \left(1+z_{coll}\right)\ .
\ee
\subsection{Closed Background}
We will now extend the above calculation to the case, where the collapsing
region resides in a curved background Universe following
\cite{Peebles1980} and \cite{Lacey:1993iv}. We will consider the case of an
overdense, closed background Universe and note that the open background can be
treated analogously. Furthermore, we will restrict ourselves to a background
Universe without a dark energy component, such that only matter and curvature
contribute to the energy budget. This closed background Universe is parametrized
as
\begin{align}
a_B=A_B \left(1-\cos{\eta}\right) && t=B_B\left(\eta-\sin{\eta}\right)
\label{eq:backpara}
\end{align}
with $\eta\in[0,2\pi]$ and the parameters
\be
A_B=\frac{4\pi G \rho_B}{3 K_B} \ , \quad{\rm and}\quad B_B=\frac{4\pi G
\rho_B}{3 K_B^{3/2}}\ .
\ee
This curved background can now be identified with the effective curved patch
describing a long wavelength fluctuation. We study the evolution of a
collapsing spherical overdensity, expanding the parametric solutions for both
the background~\eqref{eq:backpara} and the collapsing
region~\eqref{eq:collpara} at early times. This means that we restrict ourselves
to treat the curvature of the background at linear order.
The linear density contrast then scales as 
\be
\delta=\frac{a_C^3}{a_B^3}-1=\frac{3}{5}\left(\frac{3\pi}{2
}
\right)^{2/3}\left[\left(\frac{1}{t_\text{coll}}\right)^{2/3}
-\left(\frac { 1 } { t_\Omega} \right)^{2/3}\right]t^{2/3}\ ,
\ee
where $t_\Omega=2\pi B$.
For $K\to
0$ we have $t_\Omega\to \infty$ and thus we recover the EDS result
shown above.
The overdensity for an object that collapses at
$t_\text{coll}$, linearly extrapolated to present time thus reads as
\be
\frac{\delta_\text{c}(z_{coll})}{1+z_\text{coll}}=\frac{3}{5}\left(\frac{3\pi}{2
}
\right)^{2/3}\left[1-\left(\frac { t_0}
{t_\Omega} \right)^{2/3}\left(\frac { t_{coll}}
{t_0} \right)^{2/3}\right]\ .
\ee
We can now write down the collapse time for the background Universe.
\be
t_\Omega=2\pi B_B=\frac{\pi \Omega_{m,B}}{H_0 (\Omega_{m,B}-1)^{3/2}}
=\frac{\pi \left(1-\Omega_{K,B}\right)}{H_0 (-\Omega_{K,B})^{3/2}}
\ee
Using $(t_0/t_\text{coll})^{2/3}=1+z_{coll}$ and $t_0\approx
2/(3H_0)$ the overdensity of the collapsing region can be rewritten as
\be
\frac{\delta_\text{c}(z_{coll})}{1+z_\text{coll}}=\frac{3}{5}\left(\frac{3\pi}{2
}
\right)^{2/3}\left[1+\left(\frac{2}{3\pi}
\right)^{2/3}
\frac{\Omega_{K,B}}{(1-\Omega_{K,B})^{2/3}}\frac{1}{1+z_{coll}}\right]\ .
\ee
We are now going to consider the case where the curvature of the background
Universe can be described by a long wavelength fluctuation with
present day amplitude $\delta_{l,B}$. In this case we obtain for the density
parameters
of the background Universe
\be
\Omega_{m,B}=\frac{1+\delta_{l,B}}{(1-\delta_{l,B}/3)^2}=1+\frac{5}{
3 } \delta_{l,B}\quad \Rightarrow\quad \Omega_{K,B}=-\frac{5}{3}\delta_{l,B}\ .
\ee
For the overdensity of a perturbation that collapses at $z_\text{coll}$
linearly extrapolated to the present day we obtain
\be
\delta_c(z_{coll})\approx1.686\left(1+z_{coll}\right)-\delta_{l,0}=
\delta_c(\Omega_{K,B}=0)-\delta_{l,B}\ .
\ee
This result is of course very intuitive and it allows us to explicitly verify that our
General Relativistic definition of the bias agrees with the standard Newtonian one.

\section{Perturbed Geodesic Parameters\label{app:symbols}}
In this appendix we will provide the essence of cosmological perturbation
theory and expalain the gauge choices used in this paper. Then we will quickly
review the most important formulae required for the mapping to observables
before we conclude by specialising our result for the observed overdensity to
the case where the matter distribution itself is the tracer.
\subsection{Gauge Transformations}
The most general perturbed metric for a flat Universe reads as \cite{Hu:2008hd}
\be
\derd s^2=-(1+2A)\derd t^2-2 a B_i \derd x^i \derd t+a^2
\left[(1+2D)\delta_{ij}+E_{ij}\right]\derd x^i \derd x^j\ .
\ee
We will restrict ourselves to scalar modes $B_i=B Q^{(0)}_i$ and $E_{ij}=E
Q^{(0)}_{ij}$, where $Q^{(0)}$ is the scalar eigenmode of the Laplacian. Here
we perform the scalar-vector-tensor decomposition in $k$-space, where
$Q^{(0)}=\exp{\left[i \vec k \cdot \vec x\right]}$. We consider a Universe
filled with dark matter plus dark energy and neglect anisotropic stress and
pressure perturbations.\\
A gauge transformation corresponds to a change in spatial position $x^i$ and
comoving time $a \derd \tau=\derd t$
\begin{align}
 \tilde{x}^i=x^i+L^i && \tilde{\tau}=\tau+T \label{eq:gaugetrans}
\end{align}
under such a transformation the metric perturbations transform as
\begin{align}
\tilde A =A-a\dot{T}-aH T && \tilde{D}=D-\frac{k}{3}L-aHT\\ \nonumber
\tilde{B}=B+a\dot{L}+kT&& \tilde{E}_{ij}=E_{ij}+kT
\end{align}
and the components of the energy momentum tensor transform as
\begin{align}
\tilde{\delta}=\delta+3 a H T && \tilde{v}=v+a\dot{L}\ ,
\end{align}
where $v_i=v Q^{(0)}_i$. Gauge invariance refers to the fact that certain
combinations of metric and energy momentum perturbations are invariant under a
change of coordinates \eqref{eq:gaugetrans}, i.e., the numerical value of a
quantity does not change. Gauge invariance is necessary, but not sufficient, for
observability.
\par
We will consider two gauges
\begin{enumerate}
 \item Newtonian Gauge\\
Newtonian gauge is defined by $B=E=0$ setting $A=\Phi$ and $D=-\Psi$.
\be
\derd s^2=-(1+2\Phi)\derd t^2+a^2(1-2\Psi)\derd \vec x^2
\ee
Neglecting anisotropic stress we have $\Phi=\Psi$. For the Einstein equations we
have
\begin{align}
-k^2\Phi-3a^2 H^2\left(\Phi+\frac{\dot \Phi}{H}\right)=&4\pi G a^2\bar{\rho}
\delta^{(N)}\\
a H\left(\Phi+\frac{\dot \Phi}{H}\right)=&4\pi G
a^2\left(\bar{\rho}+\bar{p}\right) \frac{v^{(N)}}{k}=-a^2 \dot{H}
\frac{v^{(N)}}{k}
\end{align}
 \item Comoving Gauge\\
In comoving gauge we set $E=0$ and $\delta T^{0}_i=0$ corresponding to
$v_i=B_i$. Setting $A=\xi$ and $D=\zeta$ leads to the metric
\be
\derd s^2=-(1+2\xi)\derd t^2+a v_i \derd x^i \derd t+a^2(t)(1+2\zeta)\derd\vec
x^2
\ee
then the Einstein equations read as
\begin{align}
k^2\left(\zeta+aH\frac{v^{(com)}}{k}\right)=&4\pi G a^2
\bar{\rho}_m \delta^{(com)}\\
H\xi-\dot \zeta=&0\ ,
\end{align}
where $v^{(com)}=v^{(N)}$.
One can show that on scales larger than the sound horizon 
$\zeta$ is constant. 
Using $\dot{\zeta}=0$ we see that the lapse function $\xi$ vanishes in
pressureless media and thus the comoving gauge
is also synchronous, i.e., proper time agrees with the
coordinate time.
Using
\be
T_{{N}\to{com}}=\frac{v^{(N)}-B^{(N)}}{k}=\frac{v^{(N)}}{k}
\ee
the overdensities in the Newtonian and comoving gauge are related by
\be
\delta^{(com)}=\delta^{(N)}-a\dot{\bar{\rho}}T_{{N}\to{com}}
=\delta^{(N)}+3aH \frac{v^{(N)}}{k}\label{eq:densnewcom}
\ee
For the spatial metric perturbations we have
\be
\zeta=-\Phi-aH T_{{N}\to{com}}=-\Phi-a H
\frac{v^{(N)}}{k}=-\Phi+\frac{H^2}{\dot{H}}\left(\Phi+\frac{\dot{\Phi}}{H}
\right)
\ee
In synchronous gauge $A=B=0$ there are no sources in the equation of motion
for the velocity of stress free matter, i.e., if it was at rest initially it
will remain so for all times~\cite{Ma:1995ey}. In this case the density
perturbation in comoving and synchronous gauge agree
$\delta^{(com)}=\delta^{(syn)}$. The CMBFAST Boltzmann code \cite{Seljak:1996is} is providing the
synchronous gauge transfer function and can thus be used to infer the transfer
function for the comoving gauge density perturbation.
\end{enumerate}
Combining the Einstein equations in comoving and Newtonian gauge we have
\be
-k^2 \Phi=4\pi G a^2 \delta^{(com)}\ ,
\ee
which is valid on \emph{all} scales.
\subsection{Volume Distortion \& Observed Redshifts}
Here we explain the symbols used in eq.~(\ref{eq:observation}) for the reader's
convenience (see~\cite{Yoo:2009au} for a detailed explanation and derivation):
\begin{align}\nonumber
{\cal J}=\delta V/V=&-\Phi+v^i e_i-(1+z)\frac{d}{d z}\delta
z_{G\to z}-2\frac{1+z}{H
r}\delta z_{G\to z}-\delta z_{G\to z}\\
-&2 \kappa+\frac{1+z}{H}\frac{d
H}{dz}\delta z_{G\to z}+2 \frac{\delta
r}{r}\\ 
=&-\Phi+\left[\frac{\derd \ln H}{\derd
\ln (1+z)}-1-2\frac{1+z}{Hr_s}\right]\left[v^i e_i-\Phi-2\int_0^{r_s} \derd r\,
a\dot{\Phi} \right]\nonumber\\
+&\frac{4}{r_s}\int_0^{r_s} \derd r\, \Phi-2\int_0^{r_s} \derd r\,
\frac{r_s-r}{r
r_s}\hat{\nabla} \Phi+\frac{1}{H}\left[\dot{\Phi}-\frac{1}{a}\frac{\partial v_i
e^i}{\partial r}\right]\ . \nonumber
\end{align}
In the evaluation of the above expression we used that the total derivative
is given by $d/d z=H^{-1}d/d r=-(\partial_o-e^i\partial_i)
=-(\partial_0-\partial_r)$ and that the velocity follows the evolution equation
\be
\dot{v}_i+ H v_i = -\frac{\Phi_{,i}}{a}\ .
\ee
The perturbation to the redshift of the
source $\delta z_{G\to z}$, is given by the relationship
\be\label{eq:zeta_perturbation}
\delta z_{G\to z}=a H\delta\tau_o+\left[v_i e^i-\Phi\right]^s_0-2\int_0^{r_s}
a\dot\Phi\;
dr\ .
\ee
Here the four velocity of the source is given by
$u^\alpha=a^{-1}\bigl((1+\Phi),v^i\bigr)$ and $e^i$ is the photon propagation
direction
as seen from the observer, $r$ is the comoving line-of-sight distance, $r_s$ is
the comoving line-of-sight distance of the source. $\delta \tau_o$~is the
perturbation to
the conformal time at the time of observation, and it is just a monopole term
that cancels when measuring fluctuations. $\delta r$ is the radial displacement,
given by
\be
\delta r=\delta \tau_0+2 \int_0^{r_s}dr\; \Phi \ .
\ee
The deflection of the photons on their way form the source galaxy to the
observer can be quantified as
\bea\label{eq:PerturbedAngles}
\delta\theta&=&2\int_0^{r_s} dr \left(\frac{r_s-r}{r\,
r_s}\right)\Phi_{,\,\theta}\ , \\ \nonumber
\delta\phi&=&2\int_0^{r_s} dr\left(\frac{r_s-r}{r\,
r_s\sin \theta}\right)\Phi_{,\,\phi}\ .
\eea
The latter can be combined to calculate the distortion of the solid
angle as quantified by the convergence $\kappa$
\be
\kappa=2\int_0^{r_s}dr\; \left(\frac{r_s-r}{ r\, r_s}\right)\hat\nabla^2\Phi\ ,
\ee
where $\hat\nabla$ is the differential operator on the two dimensional unit sphere.
In case the survey is not volume limited but rather flux limited, we need to
replace ${\cal J}$ with
\be
{\cal J}\quad\rightarrow \quad{\cal J}-5p\,\delta{\cal D}_L\ ,
\ee
where $\delta{\cal D}_L$ is the perturbation to the luminosity distance, which is given by
\bea\nonumber
\frac{\delta{\cal D}_L(z)}{D_L(z)}&=&1+v_i e^i-\Phi_s-\frac{1+z_s}{H_s
r_s}\delta z_{G\to z}+\left({\cal H}_o+\frac{1}{r_s}\right)\delta \tau_o\\
&&+2\int_0^{r_s}dr\left[\frac{\Phi}{r_s}-\frac{r}{r_s}a\dot\Phi+
\frac{(r_s-r)r}{2 r_s}\left(\nabla^2\Phi-a\frac{d(a\dot
\Phi)}{dt}+2a\dot\Phi,_i e^i\right)\right]\ ,
\eea
where $D_L(z)=(1+z)r(z)$ is the unperturbed luminosity distance 
and $p$ is the slope of the luminosity function.

Using $dz=-H(1+z)dt$ we can write in a general and in Newtonian gauge
\begin{align}
\delta z_{G\to
L}=&\frac{z_G(t_L)-z_G(t_G)}{1+z_G(t_G)}=-H(t_G)(t_L-t_G)=-H\int_0^{t_G}dt
A\nonumber\\
=&-H\int_0^{t_G}dt\ \Phi=-aH \frac{v^{(N)}}{k}\ , \\
\delta z_{G\to z}=&\frac{z-z_G(t_G)}{1+z_G}
=\left[\left(v_i-B_i\right)e^i-A\right]_o^s-\int_0^{r_s} dr\ \left[a
\left(\dot{A}-\dot{D}\right)-\left(B_{i,j}+a\dot{E}_{ij}\right)e^i
e^j\right]\nonumber\\
=&\left[v_ie^i-\Phi\right]_o^s-2\int_0^{r_s}dr\, a \dot{\Phi}\ .
\end{align}
Under a gauge transformation \eqref{eq:gaugetrans} we have
\begin{align}
\widetilde{\delta z}_{G\to L}=\delta z_{G\to L}+aHT\ , && \widetilde{\delta
z}_{G\to
z}=\delta z_{G\to z}+aHT\ .
\end{align}
Thus $\delta z_{G\to L}-\delta z_{G\to z}$ does not change under the gauge
transformation and is gauge invariant.\\
For the evaluation of the full expression in eq.~\eqref{eq:finalnongauss} we
first transform all the quantities to $k$-space.
First, we consider the line of sight projection of the
velocity
\be
v_i e^i=\int \dkc\ \bigl(-i \mu v^{(N)}(\vec k)\bigr) \ Q^{(0)}(\vec k)\ ,
\ee
where $\mu=\vec x \cdot \vec k/(x k)$ is the cosine between the $k$-mode and
the line of sight. For the redshift space distortion term we have then
\be
\partial_r v_i e^i=n^j\partial_j e^i v_i=\int \dkc\ \bigl(\mu^2 v^{(N)}(\vec
k)\bigr)\ Q^{(0)}(\vec k)\ .
\ee
Thus the volume distortion term reads as
\be
\mathcal{J}(\vec k)=-\Phi(\vec k)+\mathcal{A}(z)\bigl(-i\mu
v^{(N)}(\vec k)-\Phi(\vec k)\bigr)+\bigl(f(z)-1\bigr)\Phi(\vec k)-\mu^2 
\frac{k}{aH}v^{(N)}(\vec k) \ .
\ee
The full observed density perturbation is the sum of the latter and the
perturbation in the
proper number density of tracers 
\be
\delta_p(\vec
k)=b\, \delta^{(com)}(\vec k)+b_\zeta \zeta(\vec k)
+\mathcal{B}(z)\left(-\frac{aH}{k}v^{(N)}(\vec k)-i\mu
v^{(N)}(\vec k)-\Phi(\vec k)\right)\ ,
\ee
\begin{align}\nonumber
\delta_{obs}(\vec
k)=&\Biggl\{\biggl[-2-\mathcal{A}(z)+f(z)+f(z)\frac{\mathcal{B}(z)}{
\beta(z) } +B(z)-b_{\zeta}\left(1-\frac{f(z)}{\beta(z)}\right)
\\+&\left(\mu^2\frac { f(z) } { \beta(z) } -\frac { b} { \alpha(z) }
\right)\left(\frac{k}{aH}\right)^2\biggr]
+i\left[\bigl(\mathcal{A}(z)-\mathcal{B}(z)\bigr)\mu\frac{
f(z) } { \beta(z) }
\frac{k}{aH}\right]\Biggr\}\Phi(\vec k)\ ,
\label{eq:deltaobsk}
\end{align}
where we related the density and velocity perturbations to the Newtonian gauge
metric perturbation
\begin{align}
 \delta^{(com)}(\vec k)=-\left(\frac{k}{aH}\right)^2\frac{H^2}{4\pi G
\bar{\rho}}\Phi(\vec k)\ , && v(\vec
k)=f\frac{k}{aH}\frac{H^2}{\dot{H}}\Phi(\vec
k)\ ,
\end{align}
and introduced the auxiliary functions
\be
\mathcal{A}(z)=\frac{\derd \log H}{\derd
\log
(1+z)}-1-2\frac{1+z}{Hr_s}=\frac{3}{2}\frac{(1+z)^2}{\Omega_{m,0}(1+z)^3+\Omega_
{\Lambda,0}
} -1-2\frac{(1+z)c}{r_sH}\ ,
\ee
\begin{align}
 \beta=\frac{\dot{H}}{H^2}\ , && \alpha=\frac{4\pi G \bar{\rho}}{H^2}\ , &&
\mathcal{B}(z)=\frac{\partial \log \bar{n}_p}{\partial \log
(1+z)}\ . \nonumber
\end{align}
The power spectrum is then given by
$(2\pi)^3\delta^{(D)}(\vec k+\vec k')P_{obs}(k)=\left\langle\delta_{obs}(\vec
k)\delta_{obs}^*(\vec k')\right\rangle$.

\subsection{Matter as Tracer}
For matter we have $n_p(t_L;\Omega_K)=\bar{\rho}(t_L)\bigl(1+D(t_L)
\delta_{0}^{(com)}\bigr)$. This can be seen in two ways: firstly, in
synchronous slicings the proper time and the coordinate time agree, thus the
matter overdensity in the local frame must agree with the matter overdensity in
comoving gauge. Also, we saw above in eq.~\eqref{eq:rholocal}, that the local
matter density is related to the global one by
$\bar{\rho}_L=\bar{\rho}_G(1+D(t_L)\delta_{com,0})$. Using that
$\Omega_{K,0}\propto\delta_{com,0}$
and that $\partial \log \bar{\rho}/\partial \log(1+z)=3$ we have
\be
\delta_{obs}(z,\theta,\phi)=\delta^{(com)}(z_G)+3\delta z_{G\to L}-3\delta
z_{G\to z}+{\cal{J}}\ ,
\ee
where the $\delta z$'s are in a general, yet unspecified gauge.\\
For the transformation from a comoving to general gauge we have for the
densities
\be
\delta^{(gen)}=\delta^{(com)}+3aHT_{com\to gen}\ . \label{eq:denstransgencom}
\ee
The integral entering into $\delta z_{G\to L}$ transforms as
\be
3\delta z_{G\to L}^{(gen)}=-3H\int dt A^{(gen)}=-3H\int dt A^{(com)}
+3aHT_{com \to gen}=3aHT_{com \to gen}=\delta^{(gen)}-\delta^{(com)}\ ,
\ee
where we used that $A^{(com)}=0$ and solved eq.~\eqref{eq:denstransgencom} for
$T_{com \to gen}$.
Evaluating $\delta_{obs}$ in the general gauge we obtain
\be
\delta_{obs}(z,\theta,\phi)=\delta^{(com)}-3H\int dt A^{(gen)}-3\delta
z_{G\to z}^{(gen)}+\mathcal{J}=\delta^{(gen)}-3\delta z_{G\to
z}^{(gen)}+\mathcal{J},
\ee
where $\delta^{(gen)}$ and $\delta z_{G\to L}^{(gen)}$ are the matter
overdensity and
the redshift lapse in a general gauge. This expression agrees with
$\delta_{n_p}=\delta^{(gen)}-3\delta z_{G\to z}^{(gen)}$ in \cite{Yoo:2009au}.
In the follow up paper, \cite{Yoo:2010ni} used a local bias in the
matter density at the observed redshift $\delta^{(gen)}-3\delta z^{(gen)}_{G\to
z}$. This means that in contrast to our approach the bias factor $b$ is also
multplying the redshift lapse terms between global, local and observed redshift,
i.e.\ the evolution
of the sample is fixed to be $\partial \log \bar{n}_p/\partial \log (1+z)=3b$,
while for a typical quasar sample this number can vary in a much wider range
depending on the redshift distribution of the sample.
 \begingroup\raggedright\endgroup


\begin{thebibliography}{10}

\bibitem{Yoo:2009au}
  J.~Yoo, A.~L.~Fitzpatrick and M.~Zaldarriaga,
  ``A New Perspective on Galaxy Clustering as a Cosmological Probe: General
  Relativistic Effects,''
  Phys.\ Rev.\  D {\bf 80} (2009) 083514
  [arXiv:0907.0707 [astro-ph.CO]].

\bibitem{Dalal:2007cu}
  N.~Dalal, O.~Dore, D.~Huterer and A.~Shirokov,
  ``The imprints of primordial non-gaussianities on large-scale structure:
  scale dependent bias and abundance of virialized objects,''
  Phys.\ Rev.\  D {\bf 77} (2008) 123514
  [arXiv:0710.4560 [astro-ph]].



\bibitem{Slosar:2008hx}
  A.~Slosar, C.~Hirata, U.~Seljak, S.~Ho and N.~Padmanabhan,
  ``Constraints on local primordial non-Gaussianity from large scale
  structure,''
  JCAP {\bf 0808} (2008) 031
  [arXiv:0805.3580 [astro-ph]].


\bibitem{Senatore:2010wk}
  L.~Senatore and M.~Zaldarriaga,
  ``The Effective Field Theory of Multifield Inflation,''
  arXiv:1009.2093 [hep-th].


\bibitem{Baldauf:2010vn}
  T.~Baldauf, U.~Seljak and L.~Senatore,
  ``Primordial non-Gaussianity in the Bispectrum of the Halo Density Field,''
  JCAP {\bf 1104} (2011) 006
  [arXiv:1011.1513 [astro-ph.CO]].


\bibitem{MM}
    E.~Fermi, ``Sopra i fenomeni che avvengono in vicinanza di una linea oraria,''  Atti Accad. Naz. Lincei Cl. Sci. Fis. Mat.  Nat. 31,184-187, 306-309, 1922.
  F.~K.~Manasse and C.~W.~Misner,
  ``Fermi Normal Coordinates and Some Basic Concepts in Differential Geometry,''
  J.\ Math.\ Phys.\ 4, 735 (1963).
  


\bibitem{Tormen1995}
  G.~Tormen and E.~Bertschinger,
  ``Adding Long Wavelength Modes to an $N$-Body Simulation,''
  arXiv:astro-ph/9512131.


\bibitem{Cole1996}
  S.~Cole,
  ``Adding Long-Wavelength Power to N-body Simulations,''
  arXiv:astro-ph/9604046.


\bibitem{Schneider2011}
  M.~D.~Schneider, S.~Cole, C.~S.~Frenk and I.~Szapudi,
  ``Fast generation of ensembles of cosmological N-body simulations via
  mode-resampling,''
  arXiv:1103.2767 [astro-ph.CO].
  
\bibitem{Sasaki:1987ad}
  M.~Sasaki,
  ``The Magnitude - Redshift relation in a perturbed Friedmann Universe,''
  Mon.\ Not.\ Roy.\ Astron.\ Soc.\  {\bf 228}, 653-669 (1987).

\bibitem{Dodelson:2008qc}
  S.~Dodelson, F.~Schmidt and A.~Vallinotto,
  ``Universal Weak Lensing Distortion of Cosmological Correlation Functions,''
  Phys.\ Rev.\  D {\bf 78}, 043508 (2008)
  [arXiv:0806.0331 [astro-ph]].

\bibitem{Wands:2009ex}
  D.~Wands and A.~Slosar,
  ``Scale-dependent bias from primordial non-Gaussianity in general
  relativity,''
  Phys.\ Rev.\  D {\bf 79} (2009) 123507
  [arXiv:0902.1084 [astro-ph.CO]].

\bibitem{Yoo:2010ni}
  J.~Yoo,
  ``General Relativistic Description of the Observed Galaxy Power Spectrum: Do
  We Understand What We Measure?,''
  Phys.\ Rev.\  {\bf D82}, 083508 (2010).
  [arXiv:1009.3021 [astro-ph.CO]].

\bibitem{Bartolo:2010ec}
  N.~Bartolo, S.~Matarrese, A.~Riotto,
  ``Relativistic effects and primordial non-Gaussianity in the galaxy bias,''
  JCAP {\bf 1104}, 011 (2011).
  [arXiv:1011.4374 [astro-ph.CO]].

\bibitem{Komatsu:2010fb}
  E.~Komatsu {\it et al.},
  ``Seven-Year Wilkinson Microwave Anisotropy Probe (WMAP) Observations:
  Cosmological Interpretation,''
  arXiv:1001.4538 [astro-ph.CO].

\bibitem{Senatore:2009gt}
  L.~Senatore, K.~M.~Smith and M.~Zaldarriaga,
  ``Non-Gaussianities in Single Field Inflation and their Optimal Limits from
  the WMAP 5-year Data,''
  JCAP {\bf 1001} (2010) 028
  [arXiv:0905.3746 [astro-ph.CO]].

\bibitem{Creminelli:2006xe}
  P.~Creminelli, M.~A.~Luty, A.~Nicolis and L.~Senatore,
  ``Starting the Universe: Stable violation of the null energy condition and
  non-standard cosmologies,''
  JHEP {\bf 0612} (2006) 080
  [arXiv:hep-th/0606090].
  P.~Creminelli, G.~D'Amico, J.~Norena and F.~Vernizzi,
  ``The Effective Theory of Quintessence: the $w<-1$ Side Unveiled,''
  JCAP {\bf 0902} (2009) 018
  [arXiv:0811.0827 [astro-ph]].
  P.~Creminelli, G.~D'Amico, J.~Norena, L.~Senatore and F.~Vernizzi,
  ``Spherical collapse in quintessence models with zero speed of sound,''
  JCAP {\bf 1003} (2010) 027
  [arXiv:0911.2701 [astro-ph.CO]].

  
\bibitem{Chisari:2011iq}
  N.~E.~Chisari and M.~Zaldarriaga,
  ``On the connection between Newtonian simulations and General Relativity,''
  arXiv:1101.3555 [astro-ph.CO].

\bibitem{Fitzpatrick:2009ci}
  A.~L.~Fitzpatrick, L.~Senatore and M.~Zaldarriaga,
  ``Contributions to the Dark Matter 3-Pt Function from the Radiation Era,''
  JCAP {\bf 1005} (2010) 004
  [arXiv:0902.2814 [astro-ph.CO]].

\bibitem{Cheung:2007st}
  C.~Cheung, P.~Creminelli, A.~L.~Fitzpatrick, J.~Kaplan and L.~Senatore,
  ``The Effective Field Theory of Inflation,''
  JHEP {\bf 0803} (2008) 014
  [arXiv:0709.0293 [hep-th]].

  
\bibitem{Tormen:1995sd}
  G.~Tormen and E.~Bertschinger,
  ``Adding Long Wavelength Modes to an $N$-Body Simulation,''
  arXiv:astro-ph/9512131.

\bibitem{Cole:1996hb}
  S.~Cole,
  ``Adding Long-Wavelength Power to N-body Simulations,''
  arXiv:astro-ph/9604046.

\bibitem{Seljak:1996is}
  U.~Seljak and M.~Zaldarriaga,
  ``A Line of Sight Approach to Cosmic Microwave Background Anisotropies,''
  Astrophys.\ J.\  {\bf 469} (1996) 437
  [arXiv:astro-ph/9603033].

\bibitem{Lewis:1999bs}
  A.~Lewis, A.~Challinor and A.~Lasenby,
  ``Efficient Computation of CMB anisotropies in closed FRW models,''
  Astrophys.\ J.\  {\bf 538} (2000) 473
  [arXiv:astro-ph/9911177].


\bibitem{Gangui:1993tt}
  A.~Gangui, F.~Lucchin, S.~Matarrese and S.~Mollerach,
  ``The Three Point Correlation Function Of The Cosmic Microwave Background In
  Inflationary Models,''
  Astrophys.\ J.\  {\bf 430} (1994) 447
  [arXiv:astro-ph/9312033].


\bibitem{Lyth:2002my}
  D.~H.~Lyth, C.~Ungarelli and D.~Wands,
  ``The primordial density perturbation in the curvaton scenario,''
  Phys.\ Rev.\  D {\bf 67}, 023503 (2003)
  [arXiv:astro-ph/0208055].

\bibitem{Zaldarriaga:2003my}
  M.~Zaldarriaga,
  ``Non-Gaussianities in models with a varying inflaton decay rate,''
  Phys.\ Rev.\  D {\bf 69}, 043508 (2004)
  [arXiv:astro-ph/0306006].


\bibitem{Creminelli:2007aq}
  P.~Creminelli and L.~Senatore,
  ``A smooth bouncing cosmology with scale invariant spectrum,''
  JCAP {\bf 0711} (2007) 010
  [arXiv:hep-th/0702165].


\bibitem{Chen:2009zp}
  X.~Chen and Y.~Wang,
  ``Quasi-Single Field Inflation and Non-Gaussianities,''
  JCAP {\bf 1004} (2010) 027
  [arXiv:0911.3380 [hep-th]].

\bibitem{Bruni2011}
  M.~Bruni, R.~Crittenden, K.~Koyama, R.~Maartens, C.~Pitrou and D.~Wands,
  ``Disentangling non-Gaussianity, bias and GR effects in the galaxy
  distribution,''
  arXiv:1106.3999 [astro-ph.CO].
  
\bibitem{Silverstein:2008sg}
  E.~Silverstein and A.~Westphal,
  ``Monodromy in the CMB: Gravity Waves and String Inflation,''
  Phys.\ Rev.\  D {\bf 78} (2008) 106003
  [arXiv:0803.3085 [hep-th]].

\bibitem{McAllister:2008hb} L.~McAllister, E.~Silverstein and A.~Westphal,
  ``Gravity Waves and Linear Inflation from Axion Monodromy,''
  arXiv:0808.0706 [hep-th].


\bibitem{Green:2009ds}
  D.~Green, B.~Horn, L.~Senatore and E.~Silverstein,
  ``Trapped Inflation,''
  Phys.\ Rev.\  D {\bf 80} (2009) 063533
  [arXiv:0902.1006 [hep-th]].
  
\bibitem{Barnaby:2009mc}
  N.~Barnaby, Z.~Huang, L.~Kofman and D.~Pogosyan,
  ``Cosmological Fluctuations from Infra-Red Cascading During Inflation,''
  Phys.\ Rev.\  D {\bf 80} (2009) 043501
  [arXiv:0902.0615 [hep-th]].


\bibitem{Flauger:2009ab}
  R.~Flauger, L.~McAllister, E.~Pajer, A.~Westphal and G.~Xu,
  ``Oscillations in the CMB from Axion Monodromy Inflation,''
  arXiv:0907.2916 [hep-th].
  R.~Flauger and E.~Pajer,
  ``Resonant Non-Gaussianity,''
  arXiv:1002.0833 [hep-th].

\bibitem{Chen:2006xjb}
  X.~Chen, R.~Easther and E.~A.~Lim,
  ``Large Non-Gaussianities in Single Field Inflation,''
  JCAP {\bf 0706} (2007) 023
  [arXiv:astro-ph/0611645].

\bibitem{Hirata:2002jy}
  C.~M.~Hirata and U.~Seljak,
  ``Analyzing weak lensing of the cosmic microwave background using the
  likelihood function,''
  Phys.\ Rev.\  D {\bf 67} (2003) 043001
  [arXiv:astro-ph/0209489].


\bibitem{Hirata:2003ka}
  C.~M.~Hirata and U.~Seljak,
  ``Reconstruction of lensing from the cosmic microwave background
  polarization,''
  Phys.\ Rev.\  D {\bf 68} (2003) 083002
  [arXiv:astro-ph/0306354].

\bibitem{Bernardeau:2001qr}
  F.~Bernardeau, S.~Colombi, E.~Gaztanaga and R.~Scoccimarro,
  ``Large-scale structure of the Universe and cosmological perturbation
  theory,''
  Phys.\ Rept.\  {\bf 367} (2002) 1
  [arXiv:astro-ph/0112551].


\bibitem{Challinor2011}
  A.~Challinor and A.~Lewis,
  ``The linear power spectrum of observed source number counts,''
  arXiv:1105.5292 [astro-ph.CO].

\bibitem{Bonvin2011}
  C.~Bonvin and R.~Durrer,
  ``What galaxy surveys really measure,''
  arXiv:1105.5280 [astro-ph.CO].

\bibitem{Jeong:2011as}
  D.~Jeong, F.~Schmidt, C.~M.~Hirata,
  ``Large-scale clustering of galaxies in general relativity,''
  [arXiv:1107.5427 [astro-ph.CO]].

\bibitem{Seery:2005wm}
  D.~Seery and J.~E.~Lidsey,
  ``Primordial non-Gaussianities in single field inflation,''
  JCAP {\bf 0506} (2005) 003
  [arXiv:astro-ph/0503692].
  X.~Chen, M.~x.~Huang, S.~Kachru and G.~Shiu,
  ``Observational signatures and non-Gaussianities of general single field
  inflation,''
  JCAP {\bf 0701} (2007) 002
  [arXiv:hep-th/0605045].
  L.~Boubekeur, P.~Creminelli, J.~Norena and F.~Vernizzi,
  ``Action approach to cosmological perturbations: the 2nd order metric in
  matter dominance,''
  JCAP {\bf 0808} (2008) 028
  [arXiv:0806.1016 [astro-ph]].


  
  
\bibitem{Peebles1980} Peebles, P.~J.~E.\ 1980.\  {\it The
large-scale structure of the Universe}.\ Princeton University Press, 1980.~435
p.\ .


\bibitem{Lacey:1993iv}
 C.~G.~Lacey and S.~Cole,
 ``Merger rates in hierarchical models of galaxy formation,''
 Mon.\ Not.\ Roy.\ Astron.\ Soc.\  {\bf 262}, 627 (1993).
  
\bibitem{Cooperstock:1998ny}
  F.~I.~Cooperstock, V.~Faraoni and D.~N.~Vollick,
  ``The influence of the cosmological expansion on local systems,''
  Astrophys.\ J.\  {\bf 503}, 61 (1998)
  [arXiv:astro-ph/9803097].

\bibitem{Hu:2008hd}
  W.~Hu,
  ``Lecture Notes on CMB Theory: From Nucleosynthesis to Recombination,''
  arXiv:0802.3688 [astro-ph].


\bibitem{Ma:1995ey}
  C.~P.~Ma and E.~Bertschinger,
  ``Cosmological perturbation theory in the synchronous and conformal Newtonian
  gauges,''
  Astrophys.\ J.\  {\bf 455}, 7 (1995)
  [arXiv:astro-ph/9506072].



\end{thebibliography}
\end{document}